 \pdfoutput=1
\documentclass[floatfix]{emulateapj}
\usepackage{epsfig,subfigure,color,amsmath,graphicx,multirow,subfigmat,ctable,bm}
\newcommand{\sperp}{{\scriptscriptstyle\perp}}
\newcommand{\spara}{{\scriptscriptstyle\parallel}}
\newcommand\bigzero{\makebox(0,0)[b]{\text{\bf \Large0}}}
\newcommand\zero{\makebox(0,0)[b]{\text{\bf \large0}}}
\newcommand{\appropto}{\mathrel{\vcenter{
  \offinterlineskip\halign{\hfil$##$\cr
    \propto\cr\noalign{\kern2pt}\sim\cr\noalign{\kern-2pt}}}}}
\bibpunct[; ]{(}{)}{;}{a}{}{,}

\begin{document}
\title{Bayesian semi-blind component separation for foreground removal in interferometric 21-cm observations}
\author{Le Zhang$^{1}$, Emory~F.~Bunn$^{7}$, Ata Karakci$^{2}$, Andrei Korotkov$^{2}$, P.M. Sutter$^{9,10,6}$, Peter T. Timbie$^{1}$, Gregory~S.~Tucker$^{2}$, Benjamin~D.~Wandelt$^{4,5,3,8}$\\
{~}\\
$^{1}$Department of Physics, University of Wisconsin, Madison, WI 53706, USA\\
$^{2}$Department of Physics, Brown University, 182 Hope Street, Providence, RI 02912, USA\\
$^{3}$ Department of Physics, 1110 W Green Street, University of Illinois at Urbana-Champaign, Urbana, IL 61801, USA\\
$^{4}$ UPMC Univ Paris 06, UMR 7095, Institut d'Astrophysique de Paris, 98 bis, boulevard Arago, 75014 Pharis, France\\
$^{5}$ CNRS, UMR 7095, Institut d'Astrophysique de Paris, 98 bis, boulevard Arago, 75014 Paris, France\\
$^{6}$ Center for Cosmology and Astro-Particle Physics, Ohio State University, Columbus, OH 43210, USA\\
$^{7}$ Physics Department, University of Richmond, Richmond, Virginia 23173, USA\\
$^{8}$ Department of Astronomy, University of Illinois at Urbana-Champaign, Urbana, IL 61801, USA\\
$^{9}$ INFN - National Institute for Nuclear Physics, via Valerio 2, I-34127
Trieste, Italy\\
$^{10}$ INAF - Osservatorio Astronomico di Trieste, via Tiepolo 11, 1-34143
Trieste, Italy
}
\thanks{Email: lzhang263@wisc.edu}

\begin{abstract}
We present in this paper a new Bayesian semi-blind approach for foreground removal in observations of the 21-cm signal with interferometers.  The technique, which we call HIEMICA (HI Expectation-Maximization Independent Component Analysis),  is an extension of the Independent Component Analysis (ICA) technique developed for two-dimensional (2D) CMB maps to three-dimensional (3D) 21-cm cosmological signals measured by interferometers. This technique provides a fully Bayesian inference of power spectra and maps and separates the foregrounds from signal based on the diversity of their power spectra. Only relying on the statistical independence of the components, this approach can jointly estimate the 3D power spectrum of the 21-cm signal and, the 2D angular power spectrum and the frequency dependence of each foreground component, without any prior assumptions about foregrounds. This approach has been tested extensively by applying it to mock data from interferometric 21-cm intensity mapping observations under idealized assumptions of instrumental effects. We also discuss the impact when the noise properties are not known completely. As a first step toward solving the 21 cm power spectrum analysis problem we compare the semi-blind HIEMICA technique with the commonly used Principal Component Analysis (PCA). Under the same idealized circumstances the proposed technique provides significantly improved recovery of the power spectrum.  This technique can be applied straightforwardly to all 21-cm interferometric observations, including epoch of reionization measurements, and can be extended to single-dish observations as well.


\end{abstract}


\maketitle

\section{Introduction}
One of the main goals of modern cosmology is to measure the large-scale structure (LSS) of the universe, which encodes crucial information about cosmological processes. Measuring these large-scale fluctuations is the primary tool for understanding cosmological and astrophysical phenomena and determining related parameters. Observations~\citep{2014A&A...571A..16P,2013ApJS..208...19H,smootetal1992} of the cosmic microwave background (CMB) precisely probe fluctuations induced by primordial matter density perturbations from the last scattering surface at a redshift of $z\sim 1100$ to constrain models of inflation~\citep{2003ApJS..148..213P,2006PhRvD..74b3502K}. Spectroscopic galaxy redshift surveys such as the 2dF Galaxy Redshift Survey~\citep{2003astro.ph..6581C}, WiggleZ Dark Energy Survey~\citep{2010MNRAS.401.1429D}, the Sloan Digital Sky Survey~\citep{York:2000gk}, the LAMOST~\citep{2009MNRAS.394.1775W} and Baryon Oscillation Spectroscopic Survey~\citep{Anderson:2012sa} have been attempting to map out the three-dimensional (3D) structure of galaxies in the universe, providing an even more precise measurement of cosmological parameters.   

Neutral hydrogen (HI) tomography may provide a powerful alternative to galaxy surveys. The redshifted 21-cm hyperfine transition line allows a direct measurement of the redshift (distance) of a source. Using HI tomography through the redshifted 21-cm emissions could be a relatively inexpensive tool to probe LSS over enormous volumes of our universe, from redshift 0 to ultimately as high as $\sim200$, potentially collecting orders of magnitude more information than either the CMB or galaxy surveys can provide. Cosmological parameters could be determined with far greater precision~\citep{Mao:2008ug}, and the properties of dark matter, dark energy, neutrinos~\citep{2012RPPh...75h6901P, 2010ARA&A..48..127M, 2008PhRvL.100p1301L}, and non-Gaussianity~\citep{Maldacena:2002vr,2004PhRvL..92u1301L,2014arXiv1410.7794X} from inflation could be tightly constrained.  21-cm emission may be the only way to explore the dark ages and the epoch of reionization~\citep{Madau:1996cs,Chen:2003gc,2006ApJ...653..815M,Furlanetto:2006jb}.

In particular, it is possible to trace the dynamics of the expansion of the universe and study the evolution of the dark energy by mapping the large scale structure of the universe in three dimensions~\citep{2003ApJ...598..720S}. Several authors~\citep{2010ARA&A..48..127M, Chang2008,wyithe08,2008PhRvL.100p1301L,2006astro.ph..6104P} have proposed a new technique, the so-called ``21-cm intensity mapping'', for the observation of the baryon acoustic oscillations (BAO) in the large scale structure. By mapping the 21-cm intensity field with an angular resolution of $\sim10$ arcmin across the redshift range from 0 up to 3, it may be possible to determine precisely the equation of state of the dark energy~\citep{2014arXiv1405.1452B}. The technique has been tested with existing telescopes such as the Green Bank Telescope (GBT)~\citep{2010Natur.466..463C,Masui2012,Switzer:2013ewa} and the Parkes Radio Telescope~\citep{2009MNRAS.394L...6P}. However, to obtain the sensitivity and angular resolution necessary to resolve the BAO peaks demands dedicated radio telescopes with large numbers of receivers and apertures of order 100 m.   Dedicated single-aperture telescopes (e.g. FAST~\citep{2011IJMPD..20..989N}) are prohibitively expensive.  Instead,  interferometers provide an economical way to observe large volumes of the sky and measure the BAO features efficiently.  Additionally, interferometry is well-suited to the EoR and BAO measurements: visibilities naturally measure Fourier modes of the intensity distribution on sky, which is directly related to the power spectrum of statistical fluctuations in the HI distribution. Radio interferometric arrays are already being designed and constructed. Currently, there are several dedicated HI interferometers that plan to detect the redshifted 21-cm EoR signal: LOFAR\footnote{http://www.lofar.org}, MWA\footnote{http://www.mwatelescope.org}, 21-cmA\footnote{http://21-cma.bao.ac.cn} and PAPER\footnote{http://astro.berkeley.edu/~dbacker/eor/}. GMRT\footnote{http://gmrt.ncra.tifr.res.in} is also currently under way to detect the 21 cm signal from reionization. Furthermore, several 21-cm BAO experiments have been proposed (including BAOBAB~\citep{2013ApJ...768L..36P}, BAORadio~\citep{2012CRPhy..13...46A}) or constructed (CHIME\footnote{http://chime.phas.ubc.ca}, and TianLai~\citep{2012IJMPS..12..256C}) to conduct low redshift surveys. Next-generation experiments such as SKA\footnote{http://www.skatelescope.org} will enable revolutionary progress in 21-cm tomography.

However, HI tomography is extremely challenging because astrophysical foreground contamination is expected to be at least four orders of magnitude brighter than the 21-cm signal. The success of statistical detection of the signal relies on robust foreground removal, which has been extensively explored over the past decade. Different methods have different advantages and biases in providing signal estimates. Fortunately, foregrounds are expected to have smooth continuum spectra that vary slowly with frequency, while the 21-cm signal is expected to oscillate dramatically over a short frequency range and become uncorrelated beyond frequency separations of about 1 MHz~\citep{2005ApJ...625..575S}. In principle, one can remove slowly varying components to extract the HI signal from the contaminants although this process will always leave some residual contamination.  The early ideas use a multi-frequency angular cross-correlation power spectrum to separate foregrounds from the 21-cm signal~\citep{2002ApJ...564..576D,2004MNRAS.355.1053D,2003MNRAS.346..871O,2004ApJ...608..622Z,2005ApJ...625..575S,2004ApJ...606L...5C}. An alternative approach is to subtract foregrounds along each line of sight in the sky or in the $uv$-plane, by fitting the frequency dependence with a particular smooth function (e.g. a polynomial), and then using the residuals to determine the 21-cm power spectrum~\citep{2005ApJ...625..575S,2006ApJ...650..529W,2006ApJ...653..815M,2006ApJ...638...20B,2008MNRAS.389.1319J,2008MNRAS.391..383G,2009MNRAS.398..401L,2009MNRAS.397.1138H,2011MNRAS.413.2103P,2009ApJ...695..183B,2010MNRAS.405.2492H}. Higher-order statistics, like skewness, are also employed to clean foregrounds~\citep{2009MNRAS.393.1449H}. In addition,~\cite{2009MNRAS.394.1575L,2012ApJ...757..101T,2006ApJ...648..767M} examined the impact of point-source subtraction residuals on 21-cm fluctuations, and~\cite{2004ApJ...615....7M} found a removal approach based on some symmetries in the 3D Fourier representation of radio interferometric data. More recently, simulations and observations have shown that even if sky foregrounds are spectrally smooth, the chromatic synthesized beam of an interferometer interacting with smooth spectrum foregrounds can still imprint unsmooth spectral features on measured foregrounds, contaminating a ``wedge''-like region (so-called ``mode-mixing") in cylindrical $(k_\perp,k_\spara)$ Fourier space~\citep{2010ApJ...724..526D,2012ApJ...745..176V,2012ApJ...752..137M,2012ApJ...756..165P,2012ApJ...757..101T,2013ApJ...768L..36P,2013AJ....145...65P,2013PhRvD..87d3005D,2014PhRvD..90b3018L,2014PhRvD..90b3019L,2014ApJ...782...66P}.

In order to subtract foregrounds with minimal prior assumptions on the angular distribution or frequency dependence of foreground components, some {\it blind} (non-parametric) methods have been extensively studied in the literature. By using only the data, those methods can automatically find the correlated features and separate them out into distinct components.  A commonly used data analysis technique known as Principal Component Analysis (PCA) has been applied to foreground removal and successfully tested on simulated data and real observations~\citep{2008MNRAS.388..247D,2010Natur.466..463C,Masui2012,Switzer:2013ewa}. Other non-parametric methods, including the FASTICA algorithm~\citep{fastica99,fastica10,2012MNRAS.423.2518C,2013arXiv1310.8144W} based on independent component analysis and singular value decomposition (SVD) together with an analysis based on the Karhunen-Lo\`{e}ve transform~\citep{2014ApJ...781...57S}, provide encouraging results. The FASTICA algorithm is based on the maximization of non-Gaussianity (neg-entropy), a measure of the deviation of a mixture of signals from a Gaussian distribution. However, the disadvantage of this approach is that the noise leakage into the reconstructed signal is in principle hard to estimate since the signal and noise essentially have the same statistical distribution, which is completely uncorrelated across frequencies. Other methods such as the Correlated Component Analysis (CCA)~\citep{2014arXiv1409.5300B} can be referred to as ``semi-blind'' as they require some prior knowledge. By using second-order statistics CCA also provides a promising way to separate the 21-cm reionization signal from foreground contamination for SKA measurements. Recently ~\cite{2014arXiv1409.0858V} have provided a unified approach for semi-blind foreground cleaning from multi-frequency CMB data.

In this study, we describe HIEMICA, the HI Expectation-Maximization Independent Component Analysis algorithm, a Bayesian framework to infer power spectra and maps of the HI signal and foregrounds from ``dirty'' data cubes. This technique extends the Spectral Matching ICA (SMICA)~\citep{2008arXiv0803.1814C,2003MNRAS.346.1089D,2014A&A...571A..12P} approach from two-dimensional (2D) CMB maps to the three-dimensional (3D) 21-cm cosmological signal and can be considered a semi-blind source separation method. The statistical isotropy of the cosmological Gaussian random fields in 3D Fourier space provides a characteristic and unique signature to distinguish them from the statistically non-isotropic foregrounds;  in the frequency (redshift, or radial) direction, astrophysical foreground sources contain only spectral information, which has no connection with their spatial distributions in the transverse direction.  

The purpose of this work is to take a first step toward building a new framework for 21cm data analysis. Our paper provides a complete description of the mathematical formalism and numerical techniques, which in principle can be applied to any realistic array design. Here we test the approach for an idealized array;  we propose to dedicate a future paper to systematic tests of arrays with realistic instrumental effects.

This paper is organized as follows:  Sect.~\ref{sect:ICA} describes the HIEMICA semi-blind component separation algorithm;  Sect.~\ref{sect:sim} summarizes details of our performed simulations;  Sect.~\ref{sect:app} presents the application of HIEMICA to simulated sky maps and shows the main results. Finally, Sect.~\ref{sect:con} provides discussion and concluding remarks.

\section{The HIEMICA method}\label{sect:ICA}

\subsection{Radio interferometric measurements}
For a multichannel interferometric array, the fundamental observable is the {\it visibility}, $V({\bf u}, \nu)$, which is the angular Fourier mode of the intensity fluctuations on the sky measured by a baseline ${\bf u}$ in a frequency channel $\nu$. In the flat-sky approximation, the complex visibility can be written simply as the 2D Fourier transform of the beam-modulated sky intensity:
\begin{equation}
\label{eq:vis}
V({\bf u},\nu) = \int d^2 \bm\theta\, A(\bm \theta,\nu)T(\bm \theta,\nu) e^{-2\pi i {\bf u}\cdot \bm \theta}\, ,
\end{equation}  
where  $\bm \theta$ denotes the angular sky position, $T(\bm \theta,\nu)$ is the sky temperature, $A(\bm \theta,\nu)$ is the primary beam pattern, which determines the observed area of the sky, and the baseline vector ${\bf u}$ is on the $uv$-plane perpendicular to the direction of the incoming signals, and measures the separation between the two antenna pairs in units of wavelength, i.e., ${\bf u}= \nu/c\,\Delta {\bf r}$. As the physical separation $\Delta {\bf r}$ of each pair of antennas is assumed to be fixed, each frequency channel thus gives an independent set of visibility samplings in the $uv$-plane with frequency-dependent baseline vectors.

In practice, especially for 21-cm tomography, one discretizes the 3D sky into voxels so as to efficiently estimate the intensity distribution $T(\bm \theta,\nu)$ and associated power spectrum using FFT-based algorithms. If we assume $n_\sperp$ pixels per 2D image (real or Fourier space) with even spacings in the $\theta_x$- and $\theta_y$-directions (i.e. two transverse directions), and an instrument with $n_\spara$ frequency channels of an equal spectral resolution $\Delta\nu$ to probe the sky signal in a radial direction, then we have a 3D uniform Cartesian grid of size $n_\sperp n_\spara$. Throughtout the paper, the symbols $\perp$ and $\parallel$ denote the two transverse directions and the radial direction, respectively.

Following~\cite{2012ApJS..202....9S}, the radio interferometric measurement equation of Eq.~\ref{eq:vis} at the frequency $\nu_i$, for $i=1,2,\cdots,n_\spara$, can be recast in a discrete operator formalism by  
\begin{equation}
\label{eq:visv-opt}
{\bf y_{\sperp}}[\nu_i] = {\bf I}[\nu_i] {\bf F_\sperp A}[\nu_i] {\bf F^{-1}_\sperp}  \tilde{\bm\psi }_{\sperp}[\nu_i]  + {\bf n}_{\sperp}[\nu_i], 
\end{equation}
where ${\bf y}_{\sperp}[\nu_i] = (y[u_1,\nu_i],\cdots,y[u_{n_\sperp},\nu_i])^T$ and $\tilde{\bm\psi}_{\sperp}[\nu_i] = (\tilde{\psi}[u_1,\nu_i],\cdots,\tilde{\psi}[u_{n_\sperp},\nu_i])^T$ denote, respectively,  the discretized visibilities in the $uv$-plane and the 2D Fourier ``image'' of the true angular sky ($\bm\psi_{\sperp}[\nu_i]$) of 21-cm and foregrounds at the $i$-th channel. The vector ${\bf n}_{\sperp}[\nu_i]=(n[u_1,\nu_i],\cdots,n[u_{n_\sperp},\nu_i])^T$ denotes the instrumental noise at the $i$-th channel for all the $uv$ points. Hereafter, bold, upper and lower case letters denote a matrix and a vector, respectively, and the symbol $\widetilde{~~}$ is used for Fourier components. ${\bf F_\sperp}$ represents the 2D Fourier transform operator performed in the transverse direction, converting from the real-space domain into the $uv$-space domain (of course, ${\bf F_\sperp^{-1}}$ is its inverse), and ${\bf A}[\nu_i]$ represents the primary beam pattern of the interferometer measured at the frequency $\nu_i$ and implicitly is a $n_\sperp\times n_\sperp$ diagonal matrix. The operator ${\bf I}[\nu_i]$  specifies the visibility sampling function determined by the distribution of baselines and can also be represented by a $n_\sperp\times n_\sperp$ diagonal matrix that has 1's and 0's on the diagonal to indicate whether a given pixel in $uv$-space has been observed or not.

It is convenient to introduce a vector form for numerical operations. By stacking each vector $\tilde{\bm\psi}_{\sperp}[\nu_i]$ as $\tilde{\bm \psi}_\sperp=\left[ \tilde{\bm\psi}^T_{\sperp}[\nu_1],\cdots, \tilde{\bm\psi}^T_{\sperp}[\nu_{n_\spara}]\right]^T$, Eq.~\ref{eq:visv-opt} can account for all the frequency channels when rewritten as  
\begin{eqnarray}
\label{eq:vis-opt}
{\bf y}&=& {\bf \Phi} \tilde{\bm \psi}_\sperp  + {\bf n}\,, \quad {\bf \Phi} =  \begin{pmatrix} {\bf \Phi}_1 &&\zero \\ &\ddots& \\ \zero & & {\bf \Phi}_{n_\spara} \end{pmatrix}\,, \\
{\rm with}\quad {\bf \Phi}_i &=& {\bf I}[\nu_i]{\bf F_\sperp A}[\nu_i]{\bf F_\sperp^{-1}}\,, \nonumber 
\end{eqnarray} 
where $\tilde{\bm\psi}_\sperp$ is a vector with length $n_\sperp n_\spara$ and ${\bf \Phi}$ is a block diagonal matrix, while ${\bf y}= \left[{\bf y}^T_{\sperp}[\nu_1],\cdots, {\bf y}^T_{\sperp}[\nu_{n_\spara}]\right]^T$, and ${\bf n}= \left[{\bf n}^T_{\sperp}[\nu_1],\cdots, {\bf n}^T_{\sperp}[\nu_{n_\spara}]\right]^T$, each collect vectors for all frequencies into an $n_\sperp n_\spara$-element vector. 

Notice that Eq.~\ref{eq:vis-opt} is presented on a uniform grid in order to enable the FFT operation.  Realistic observations from radio interferometers do not sample visibility data on a uniform grid in $uv$-space. Therefore, analyzing the data on a uniform grid requires an additional process, ``{\it gridding}'', to map raw visibilities from specified $uv$ points to grids with even spacing. However, the gridding process can distort the signal estimation, especially for small 21-cm signals. Alternatively,
one can use the Non Uniform Discrete Fourier Transform (NuDFT)~\citep{Fessler03nonuniformfast} to directly calculate visibility data at continuous coordinates so as to avoid ``gridding'' effects.  We leave an investigation of the gridding effects and a possible solution by NuDFT to future work.

\subsection{Linear mixture model}
The 21-cm signal along the line-of-sight for each individual pixel is not a smooth function of frequency, while most astrophysical contaminants should have smooth power-law spectral structures. If we assume that the total foreground emission contributing to the observed data in a given frequency channel can be expressed as a linear superposition of a few components and the spectral behavior of each component does not vary across the sky, the true sky temperature in pixel $\theta_k$, for $k=1,2,\cdots,n_\sperp$, in the $i$-th channel then can be modelled as 
\begin{equation}\label{eq:modelaij}
\psi[\nu_i,\theta_k] = s[\nu_i,\theta_k]+ \sum_{j=1}^{n_c} M_{i j}f_j[\theta_k] \, ,
\end{equation}
where $\nu_i$ is the frequency, $\theta_k$ is the pixel position, $n_c$ the assumed total number of foreground components, $s[\nu_i,\theta_k]$ the 21-cm signal and $f_j[\theta_k]$ the $j$-th foreground component. $M_{i j}$ reflects the contribution of the $j$-th foreground component to the $i$-th frequency channel and is assumed to be constant and independent of $\theta_k$, and thus the matrix ${\bf M}$ is commonly called the ``{\it mixing matrix}'' with dimensions of $n_\spara\times n_c$. Each column of ${\bf M}$ encodes the frequency-dependence of each foreground component.

As the Fourier modes of the isotropic 21-cm signal $s(\nu,{\bm \theta})$ are expected to be mutually independent, expressing its temperature in the Fourier coordinates $(\eta, {\bf u})$ is particularly useful. By taking the 3D Fourier transform along the frequency axis and two angular directions, one has 

\begin{equation}
\label{eq:3dFs}
\tilde{s}(\eta,{\bf u}) = \int s(\nu, \bm \theta)\, e^{-2\pi i \,(\nu \eta+{\bf u}\cdot \bm \theta)}\, d\bm \theta d\nu \, ,
\end{equation}
where $\eta$ is the Fourier dual of the frequency variable $\nu$. Note that here we perform the ``true'' Fourier transform of the sky temperature along the line-of-sight direction that is perpendicular to the $uv$-plane, rather than a ``{\it delay transform}".  The delay transform is the Fourier transform in frequency of the spectrum measured by a single baseline, which varies with frequency (i.e., ${\bf u}$ is  a function of $\nu$ in Eq.~\ref{eq:3dFs}) and which leads to the ``mode-mixing'' phenomenon (see ~\cite{2014PhRvD..90b3018L} for details).

 The foreground components can also be expressed in terms of their 2D Fourier modes. For the $j$-th component, one has
\begin{equation}
\label{eq:2dFf}
 \tilde{f}_j({\bf u}) = \int f_j(\bm \theta)\, e^{-2\pi i \,{\bf u}\cdot \bm \theta}\, d\bm \theta ~~~~~~ (j=1,\cdots,n_c)\, .
\end{equation}

According to the above equations, by performing the angular Fourier transform on Eq.~\ref{eq:modelaij} and expressing the 21-cm signal in terms of its 3D Fourier modes, Eq.~\ref{eq:modelaij} can be rewritten in a discretized version  as
\begin{equation}\label{eq:modeluv}
\tilde{\psi}[\nu_i,u_k] = \sum_{m=1}^{n_\spara} F_\spara^{-1}[i,m] \tilde{s}[\eta_m,u_k]+ \sum_{j=1}^{n_c} M_{i j}\tilde{f}_j[u_k] \, ,
\end{equation}
where $i$ denotes the frequency channel number, $k$ denotes pixel position in the $uv$-plane and we have introduced the operator (matrix) ${\bf F_\spara^{-1}}$, with its elements given by Fourier coefficients $F_\spara^{-1}[i,m]~(i=1,\cdots,n_\spara;m=1,\cdots,n_\spara)$, to represent the 1-D inverse Fourier transform converting the $\eta$-domain to the frequency domain.

Using matrix multiplications, we can rewrite the linear system above as the matrix equation: 
\begin{equation}
\label{eq:XFM}
  \tilde{\bm\Psi}= {\bf F^{-1}_\spara} \tilde{{\bf S}} +  {\bf M} \tilde{{\bf F}}\,,
\end{equation}
where $\tilde{\psi}[\nu_i,u_k]$ is the $(i,k)$-th entry of the matrix $\tilde{\bm\Psi}$, $\tilde{s}[\eta_m,u_k]$ is the $(m,k)$-th entry of the matrix $\tilde{{\bf S}}$, and $\tilde{f}_j[u_k]$ the $(j,k)$-th entry of the matrix $ \tilde{{\bf F}}$. Applying a ``vectorization'' operator ($\textrm{vec}[\cdot]$) that converts the matrix into a column vector by stacking the columns into a long column vector, one can express $ \tilde{\bm\psi}_\sperp$ in Eq.~\ref{eq:vis-opt} with $ \tilde{\bm\Psi}^T$ by 
\begin{equation}
\label{eq:3dvx}
 \tilde{\bm\psi}_\sperp=\textrm{vec}[\tilde{\bm\Psi}^T] = \textrm{vec}[({\bf F^{-1}_\spara} \tilde{{\bf S}})^T] + \textrm{vec}[({\bf M} \tilde{{\bf F}})^T]\, .
\end{equation}
Substituting Eq.~\ref{eq:3dvx} into \ref{eq:vis-opt}, and using the identity $\textrm{vec}[{\bf PQ}] = ({\bf Q}^T\otimes {\bf I}_m)\textrm{vec}[{\bf P}]$ for any matrices ${\bf P}_{m\times n}$ and ${\bf Q}_{n\times p}$, where $\otimes$ denotes the Kronecker product and ${\bf I}_m$ is an ($m\times m$) identity matrix, yields 
\begin{eqnarray}\label{eq:modelall}
 \tilde{\bm\psi}_\sperp = \left({\bf B \otimes I}_{n_\sperp}\right) {\bf x}\,,\qquad {\bf B}= \begin{pmatrix}{\bf F^{-1}_\spara} & {\bf M}\end{pmatrix} \,,
\end{eqnarray}
where ${\bf x}=( \tilde{{\bf s}}^T,\, \tilde{{\bf f}}^T)^T$ with $ \tilde{{\bf s}}=\textrm{vec}[ \tilde{{\bf S}}^T],\, \tilde{{\bf f}}= \textrm{vec}[\tilde{{\bf F}}^T]$, ${\bf I}_{n_\sperp}$ is the $n_\sperp\times n_\sperp$ identity matrix, and ${\bf B}$ is a partitioned matrix with dimensions of $n_\spara \times (n_\spara+ n_c)$. Explicitly, $\tilde{{\bf s}}$ and $ \tilde{{\bf f}}$ are vectors of length $n_\sperp n_\spara$ and $n_\sperp n_c$, respectively.  They are constructed by stacking $\tilde{{\bf s}}[\eta_m]$ and $ \tilde{{\bf f}}_j$ as $\tilde{{\bf s}}=\left( \tilde{{\bf s}}^T[\eta_1],\cdots, \tilde{{\bf s}}^T[\eta_{n_\spara}]\right)^T$ and $ \tilde{{\bf f}}=\left( \tilde{{\bf f}}^T_1 ,\cdots, \tilde{{\bf f}}^T_{n_c}\right)^T$, where the vectors $\tilde{{\bf s}}[\eta_m]$ and $\tilde{{\bf f}}_j$, each with length $n_\sperp$,  denote an image of the 3D Fourier modes of the 21-cm signal at $\eta_m$ and an image of the 2D Fourier modes of the $j$-th foreground component, respectively, i.e., $\tilde{{\bf s}}[\eta_m]=\left( \tilde{s}[u_1,\eta_m],\cdots,  \tilde{s}[u_{n_\sperp},\eta_{m}]\right)$ and $\tilde{{\bf f}}_j=\left( \tilde{f}_j[u_1] ,\cdots,  \tilde{f}_{j}[u_{n_\sperp}]\right)$.
 
By inserting Eq.~\ref{eq:modelall} into Eq.~\ref{eq:vis-opt}, the measurement equation finally can be expressed as
\begin{equation}
\label{eq:visfinal}
{\bf y = Hx}  + {\bf n}\, , \qquad {\bf H =\Phi} \left({\bf B \otimes I}_{n_\sperp}\right).
\end{equation} 

\subsection{ICA assumption}
The Independent Component Analysis (ICA) assumption is that the data can be considered as a linear mixture of a set of statistically mutually independent components.  The cosmological 21-cm signal is expected to be well approximated by an isotropic Gaussian random field and uncorrelated with foregrounds. If we also assume that the diffuse foregrounds consist of several statistically Independent Components (ICs), each of them also being an isotropic Gaussian random field with zero mean, as defined in Eq.~\ref{eq:modelall}, the covariance matrix of ${\bf x}$ then becomes a diagonal matrix, namely
\begin{equation}\label{eq:Cov-cp}
{\bf C = \left<x x^\dagger \right>}=\begin{pmatrix}{  \left<\tilde{{\bf s}}  \tilde{{\bf s}}^\dagger\right>} && \zero  \\ \zero &&  \left<{\tilde{{\bf f}}  \tilde{{\bf f}}^\dagger}\right>\end{pmatrix} .
\end{equation}
The unknown diagonal matrices $\left<\tilde{{\bf s}}  \tilde{{\bf s}}^\dagger\right>$ and $\left<{\tilde{{\bf f}}  \tilde{{\bf f}}^\dagger}\right>$ are to be estimated from the observed data.  They determine the 3D power spectrum of the HI signal and the angular power spectra of the ICs, defined through
\begin{eqnarray}\label{eq:pku}
\left<\tilde{s}[u_k,\eta_m]\tilde{s}^*[u_{k'},\eta_{m'}]\right>&=&P_{\rm HI}(u_k,\eta_m)\delta_{kk'}\delta_{mm'}\\
\left<\tilde{f}_j(u_k)\tilde{f}_{j'}^*(u_{k'})\right>&=&\delta_{jj'}\delta_{kk'} C^j_f(\ell=2\pi |{\bf u}|) 
\end{eqnarray} 
where $P_{\rm HI}(u_k,\eta_m)$ denotes the 3D power spectrum of the 21-cm signal as a function of ${\bf u}$ and $\eta$, and $C^j_f(\ell)$ denotes the angular power spectrum of the $j$-th foreground component as a function of multipole $\ell$.  The relation $\ell =2\pi |{\bf u}|$, where ${\bf u}$ is determined by the pixel position $u_k$, has been used in the flat-sky approximation.  Notice that physical sources such as the Galactic synchrotron and free-free emissions actually do have a non-zero cross-correlation and their spatial distributions clearly appear to be non-isotropic.  Nevertheless, as we will demonstrate, the ICA assumption can be regarded as an effective decomposition of sources and does not appear to affect our ability to remove foregrounds.  

\subsubsection{Noise}
We also assume that the instrument noise is an uncorrelated Gaussian distribution and can be obtained from a reasonable noise model through  
\begin{eqnarray}\label{eq:pkun}
\left<\tilde{n}(u_k,\nu_i)\tilde{n}^*(u_{k'},\nu_{i'})\right>&=&P_{\rm N}(u_k,\nu_i)\delta_{kk'}\delta_{ii'} \, .
\end{eqnarray} 

The covariance matrix thus has the form of a known block diagonal matrix (with ${\bf n}$ defined in Eq.~\ref{eq:vis-opt}), where each block is also diagonal,  
\begin{equation}
{\bf N =\left< nn^\dagger\right>}=\begin{pmatrix} {\bf N}_1 &&\bigzero \\ &\ddots& \\ \bigzero & & {\bf N}_{n_\spara} \end{pmatrix}\,.
\end{equation}
Here ${\bf N}_i$ denotes the $n_\sperp\times n_\sperp$ covariance matrix of the noise map in the $uv$-plane at the $i$-th frequency channel, which is also diagonal.

\subsection{The EM algorithm}

The goal of our analysis is to identify and separate the components from visibility data that contains mixtures of foregrounds and signal.  Using the Bayesian framework proposed by~\cite{Snoussi:2001bw} for CMB data, we will show that, without any assumption on priors, by performing a semi-blind independent component analysis for which only the noise covariance matrix ${\bf N}$ is well known,  one can successfully separate the components by jointly estimating the covariance matrix ${\bf C}$ and the mixing matrix ${\bf M}$, and can accurately estimate the power spectrum of the HI signal.

In Bayesian inference, information about unknown parameters $\bm \theta$ that we want to estimate from the data ${\bf y}$ is expressed in the form of a posterior probability distribution. Using Bayes' theorem, it can be computed as $p(\bm\theta|{\bf y}) \propto {\cal L}({\bf y}|\bm\theta)p(\bm\theta)$ where ${\cal L}({\bf y}|\bm\theta)$ is the likelihood and $p(\bm\theta)$ is the prior distribution of $\bm \theta$;  ${\bf \bm \theta=(C, M)}$ in our case. If we assume flat priors for the mixing matrix and the power spectra of all sources, and assume they are uncorrelated and independent, the prior then reduces to $P(\bm \theta)= P({\bf C})P({\bf M})\propto1$. Thus, exploring the observed-data posterior $p({\bf y}|\bm\theta)$ is equivalent to exploring the likelihood. Following~\cite{Snoussi:2001bw}, given the data model in Eq.~\ref{eq:visfinal}, ${\bf y = Hx + n}$, the mixing matrix ${\bf M}$ and the covariance matrix ${\bf C}$ can be estimated by maximizing the likelihood function. For independent and Gaussian sources, the likelihood can be expressed as 
\begin{eqnarray}\label{eq:like}
-2\ln{{\cal L}({\bf y}|\bm \theta)}&\propto& \ln{|{\bf H C H^\dagger + N}|} \\
&&+ \textrm{Tr}\left[({\bf H C H^\dagger + N})^{-1} {\bf yy^\dagger} \right]\nonumber\,.
\end{eqnarray}
Unfortunately, in general, such a likelihood evaluation is computationally intractable when applied to large datasets with realistic interferometric observations.  However, in the analogous case of CMB imaging observations$\footnotemark[$\dagger$]$\footnotetext[$^\dagger$]{The convolution with a frequency-dependent beam response in real space becomes a simple product in Fourier space.}, this likelihood can be approximated by the SMICA-likelihood~\citep{2003MNRAS.346.1089D} to measure a ``spectral mismatch'' in the Fourier domain between the empirical covariance matrices of the data and their ensemble averages, which depend on the estimated parameters. In practice, minimization of the spectral mismatch is equivalent to maximizing the likelihood and can be achieved with the Expectation-Maximization (EM) algorithm.  The EM algorithm has been applied to estimate the CMB power spectrum and reconstruct the CMB map from multi-frequency microwave maps~\citep{2003MNRAS.346.1089D}. The EM algorithm is an elaborate technique to find the maximum-likelihood estimate of the parameters when directly evaluating the likelihood function is intractable.  The calculation is simplified by assuming the existence of additional but missing parameters. It is an iterative algorithm to repeatedly solve a tractable complete-data problem instead of solving a difficult incomplete-data problem. 

Here we briefly summarize the EM algorithm when applied to our model. The likelihood $p({\bf y}|\bm\theta)$ first can be obtained by marginalizing the joint distribution $p({\bf y, x}|\bm \theta)$ over the missing data ${\bf x}$ as 

\begin{equation}
p({\bf y}|\bm \theta) =\int p({\bf y, x}|\bm \theta)d{\bf x}\,.
\end{equation}

The key idea is that the EM algorithm does not maximize $p({\bf y}|\bm\theta)$ directly; instead, it maximizes the so-called EM-functional as follows:
\begin{eqnarray}
Q(\bm \theta|\bm \theta^{n}) &=& \text{E}\left[\ln  p({\bf y},{\bf x}|\bm\theta)|{\bf y},\bm \theta^{n}\right] \nonumber\,\\
&=&\int \ln p({\bf y},{\bf x}|\bm\theta) p({\bf x}|{\bf y},\bm\theta^n)d{\bf x}\label{eq:q}\, ,
\end{eqnarray}
which is the expected value of the complete-data log-likelihood with respect to the missing data ${\bf x}$ given the observed data ${\bf y}$ and the current parameter estimates $\bm\theta^{n}$. The evaluation of this expectation is called the \textit{Expectation (E)-step} of the algorithm.

In our model, the prior distribution of the complex Fourier modes ${\bf x}$ is assumed to be Gaussian with zero mean,
\begin{equation}
p({\bf x|C}) \propto \exp\left(-\frac{1}{2} {\bf x^\dagger C^{-1} x}\right)\, ,
\end{equation}
where the diagonal covariance matrix ${\bf C}$ is defined in Eq.~\ref{eq:Cov-cp}.  Within a Bayesian framework, the joint posterior  distribution $p({\bf y, x}|\bm \theta)$   can be expressed by $p({\bf y, x}|\bm\theta) =  p({\bf y}|{\bf x}, \bm \theta) p({\bf x|\bm \theta})$, yielding 
\begin{eqnarray}\label{eq:pyx}
-2\ln p({\bf y, x}|\bm\theta) = &&\ln|{\bf N}| +{\bf  (y-Hx)^\dagger N^{-1}(y-Hx)}  \nonumber\\
&&+\ln|{\bf C}|+{\bf  x^\dagger C^{-1}x} +cst. \,
\end{eqnarray}

Using Bayes' rule, the conditional probability distribution function $p({\bf x}|{\bf y},\bm\theta^n)$ for the signal given the data is also the Gaussian,  
\begin{eqnarray}\label{eq:px|y}
p({\bf x|y},\bm\theta^n) \propto \exp\left( -\frac{1}{2}{\bf  (x-\hat{x} )^\dagger \Sigma^{-1} (x-\hat{x})}\right) \,,
\end{eqnarray}
where 
\begin{eqnarray}
&&{\bf \hat{x}}= [{\bf H}^\dagger {\bf N ^{-1}} {\bf H}+ {\bf C^{-1}} ]^{-1} {\bf H}^\dagger {\bf N^{-1}}{\bf y}\label{eq:wf} \,,\\
&&{\bf \Sigma =(H^\dagger N^{-1}H+C^{-1})^{-1}}\,, \label{eq:Sigma}
\end{eqnarray}
 where ${\bf \hat{x}}$ is the so-called Wiener-Filtered (WF) map and ${\bf \Sigma}$ is the corresponding covariance ${\bf \Sigma}= \left<({\bf x -\hat{x}})({\bf x -\hat{x}})^\dagger\right>$.   The solution for ${\bf \hat{x}}$ is the general map-making problem in cosmology. This WF map can be computed efficiently by implementing a preconditioned conjugate-gradient method that allows one to iteratively reach the solution in a tractable amount of computation time. 

Using the above Eqs.~\ref{eq:pyx} and \ref{eq:px|y}, we integrate out ${\bf x}$ to derive the expression of Eq.~\ref{eq:q} given by
\begin{eqnarray}\label{eq:Qi}
Q(\bm\theta|\bm\theta^n)=cst. -\frac{1}{2}\ln |{\bf N}| - \frac{1}{2}\ln |{\bf C}| -\frac{1}{2} \textrm{Tr}\left[{\bf C}^{-1}\widehat{{\bf R}}^\dagger_{xx} \right]&&\quad~\\
-\frac{1}{2}\textrm{Tr}\left[{\bf N}^{-1}\left(
 \widehat{{\bf R}}_{{\bf yy}} -{\bf H}\widehat{{\bf R}}^\dagger_{{\bf yx}} -\widehat{{\bf R}}_{{\bf yx}} {\bf H^\dagger } + {\bf H}\widehat{{\bf R}}_{{\bf xx}} {\bf H}^\dagger  \right) \right]&&\,\nonumber 
\end{eqnarray}
where 

\begin{eqnarray}
 {\bf  \widehat{R}_{yy}}= {\bf y y^\dagger}\,,~~   
 {\bf  \widehat{R}_{yx}}=  {\bf y \hat{x}^\dagger}\,,~~
 {\bf  \widehat{ R}_{xx}}=  {\bf  \Sigma} +  {\bf \hat{x}} {\bf \hat{x}}^\dagger\, \label{eq:cx}.  
\end{eqnarray}
Notice here ${\bf  \widehat{R}_{yy}}, {\bf  \widehat{R}_{yx}}$ and ${\bf  \widehat{ R}_{xx}}$ only depend on $\bm \theta^n$ rather than $\bm \theta$. 


The second step, called  the \textit{Maximization (M)-step}, updates the parameters by maximizing the expectation we computed in the previous E-step. These two steps are repeated as necessary. This procedure is guaranteed to increase the likelihood $p({\bf y}|\bm \theta)$ monotonically with successive iterations. In order to obtain the parameter $\bm \theta^{n+1}$ at iteration $n+1$, we solve the gradient equation with respect to ${\bf M}$ and ${\bf C}$ to maximize the functional $Q$. To do so, let us first introduce some notation that allows us to express the derivative simply.
 
Let $\bf{\hat{x}}$ and ${\bf \Sigma}$ be partitioned as follows:   
\begin{equation}
{\bf \hat{x}}= \left( \hat{\tilde{{\bf s}}}^T,\hat{\tilde{{\bf f}}}^T\right)^T\,
\end{equation}
with
\begin{eqnarray}
\hat{\tilde{{\bf s}}}= \left( \hat{\tilde{{\bf s}}}^T_\sperp[\eta_1],\cdots,\hat{\tilde{{\bf s}}}^T_\sperp[\eta_{n_\spara}]\right)^T \,, \hat{\tilde{{\bf f}}}= \left(\hat{\tilde{{\bf f}}}^T_1,\cdots,\hat{\tilde{{\bf f}}}^T_{n_c}\right)^T\,.
\end{eqnarray}
Here  ${\bf \hat{x}}$ is equally divided into $(n_\spara+n_c)$ subvectors, each with length  $n_\sperp$, ${\bf \hat{s}}_\sperp[\eta_i]$ denotes the WF map of the 21-cm signal at $\eta_i$, and ${\bf \hat{f}}_j$ denotes the WF map of the $j$-th foreground component.  Similarly 
 
\begin{equation}
{\bf \Sigma} =    \begin{pmatrix}  {\bf \Sigma}^{\tilde{s}\tilde{s}}& {\bf \Sigma}^{\tilde{s}\tilde{f}}\\
 {{\bf \Sigma}^{\tilde{s}\tilde{f}}}^\dagger& {\bf \Sigma}^{\tilde{f}\tilde{f}}      \end{pmatrix}       
\end{equation}
with
\begin{eqnarray}
 {\bf \Sigma}^{\tilde{s}\tilde{s}} &=&\begin{pmatrix} {\bf \Sigma}^{\tilde{s}\tilde{s}}_{11} &\cdots&{\bf \Sigma}^{\tilde{s}\tilde{s}}_{1n_\sperp}\\
 \vdots&\ddots&\vdots\\
{\bf \Sigma}^{\tilde{s}\tilde{s}}_{n_\sperp1} &\cdots&{\bf \Sigma}^{\tilde{s}\tilde{s}}_{n_\sperp n_\sperp}
\end{pmatrix}
~~
{\bf \Sigma}^{\tilde{s}\tilde{f}}=\begin{pmatrix}{\bf \Sigma}^{\tilde{s}\tilde{f}}_{11}&\cdots&{\bf \Sigma}^{\tilde{s}\tilde{f}}_{1n_c} \\
 \vdots&\ddots&\vdots\\ 
{\bf \Sigma}^{\tilde{s}\tilde{f}}_{n_\sperp1}&\cdots&{\bf \Sigma}^{\tilde{s}\tilde{f}}_{n_\sperp n_c}
\end{pmatrix} 
\nonumber \\
{\bf \Sigma}^{\tilde{f}\tilde{f}} &=&\begin{pmatrix}{\bf \Sigma}^{\tilde{f}\tilde{f}}_{11}&\cdots&{\bf \Sigma}^{\tilde{f}\tilde{f}}_{1n_c} \\
\vdots&\ddots&\vdots \\
{\bf \Sigma}^{\tilde{f}\tilde{f}}_{n_c1}&\cdots&{\bf \Sigma}^{\tilde{f}\tilde{f}}_{n_cn_c}
\end{pmatrix} 
\,,
\end{eqnarray}
where
\begin{eqnarray}
{\bf \Sigma}^{\tilde{s}\tilde{s}}_{ij}&=& \left<(\tilde{{\bf s}}[\eta_i]- \hat{\tilde{{\bf s}}}[\eta_i])(\tilde{{\bf s}}[\eta_j]-\hat{\tilde{{\bf s}}}[\eta_j])^\dagger\right>\,, \\
{\bf \Sigma}^{\tilde{s}\tilde{f}}_{ij}&=&\left<(\tilde{{\bf s}}[\eta_i]- \hat{\tilde{{\bf s}}}[\eta_i])(\tilde{{\bf f}}[\eta_j]-\hat{\tilde{{\bf f}}}[\eta_j])^\dagger\right>\,, \\
 {\bf \Sigma}^{\tilde{f}\tilde{f}}_{ij}&=&\left<(\tilde{{\bf f}}[\eta_i]- \hat{\tilde{{\bf f}}}[\eta_i])(\tilde{{\bf f}}[\eta_j]-\hat{\tilde{{\bf f}}}[\eta_j])^\dagger\right>\,,
\end{eqnarray}
 each with dimensions of  $n_\sperp \times n_\sperp$.

The gradient equation then reads:
\begin{eqnarray}
\frac{\partial Q} {\partial{\bf C}} &=& -({\bf C^{-1}}) + {\bf C^{-1}} {\bf \widehat{R}_{xx}}  {\bf C^{-1}}  =0~~~~~ \label{eq:Qdc}\\
\frac{\partial Q}{\partial  M_{ij}} &=& \textrm{Tr}\left[{\bf N_{\it i}^{-1}}\left( {\bm\Phi}_i \hat{\tilde{{\bf f}}}_j{\bf y}^\dagger_i + ({\bm\Phi}_i  \hat{\tilde{{\bf f}}}_j{\bf y}^\dagger_i)^\dagger\right) \right]\, \nonumber\\
&&- \frac{\partial \textrm{Tr}[{\bf N}^{-1} {\bf H}{\bf \widehat{R}_{xx}} {\bf H^\dagger }  )]}{\partial M_{ij}}=0 \label{eq:Qdmij}\, 
 \label{eq:Qdq}
\end{eqnarray} 
where the derivative of the last term can be well approximated by 
\begin{eqnarray}
&&\frac{\partial \textrm{Tr}[{\bf N}^{-1} {\bf H}{\bf \widehat{R}_{xx}} {\bf H^\dagger}  )]}{\partial M_{ij}}\approx \\
&&\sum_{k=1}^{n_c} M_{ik}\textrm{Tr}\left[{\bf N_{\it i}^{-1}} {\bm \Phi}_i\left({\bf \Sigma}^{\tilde{f}\tilde{f}}_{kj} + \hat{\tilde{{\bf f}}}_k \hat{\tilde{{\bf f}}}_j^\dagger +({\bf \Sigma}^{\tilde{f}\tilde{f}}_{kj} +  \hat{\tilde{{\bf f}}}_k \hat{\tilde{{\bf f}}}_j^\dagger)^\dagger \right){\bm\Phi^\dagger_{\it i}}\right]\nonumber\,
 \label{eq:dtrdmij}
\end{eqnarray} 
where ${\bm\Phi}_i= {\bf I}[\nu_i] {\bf F_\sperp A}[\nu_i] {\bf F_\sperp^{-1}}$ as defined in Eq.~\ref{eq:visv-opt}, and we have neglected the terms that include ${\bf \Sigma}^{\tilde{s}\tilde{f}}$ and ${\bf \Sigma}^{\tilde{s}\tilde{s}}$, although the complete expression can be obtained straightforwardly. This is an excellent approximation for 21-cm foregrounds since they are several orders of magnitude brighter than the 21-cm signal$\footnotemark[$\dagger$]$. \footnotetext[$\dagger$]{Notice that, in the case of CMB, the evaluation of the mixing matrix not only depends on CMB foregrounds but also on the CMB signal itself since they are roughly comparable.}By making this approximation, we are able to dramatically speed up the computation. According to Eqs.~\ref{eq:Qdc}, \ref{eq:Qdmij} and \ref{eq:dtrdmij}, one can establish the {\it update scheme} as follows:
\small
\begin{eqnarray}
{\bf M}_{i,:} && \leftarrow \left(\rho^{\tilde{f}y}_i[1], \cdots, \rho^{\tilde{f}y}_i[n_c]\right) \begin{pmatrix}\rho^{\tilde{f}\tilde{f}}_i[1,1] &\cdots& \rho^{\tilde{f}\tilde{f}}_i[1,n_c] \\\vdots  & \ddots &\vdots \\
 \rho^{\tilde{f}\tilde{f}}_i[n_c,1]&\cdots& \rho^{\tilde{f}\tilde{f}}_i[n_c,n_c]\end{pmatrix}^{-1} \,\\
{\bf C}_{ii}&& \leftarrow  ({\bf \widehat{R}_{xx}})_{ii}\label{eq:upcl}\,
\end{eqnarray}
\normalsize
where ${\bf M}_{i,:}$ denotes the $i$-th row of the mixing matrix, and 
\begin{eqnarray}\label{eq:psiff}
&&\rho^{\tilde{f}\tilde{f}}_i[k,j]={\rm Re}\left\{\textrm{Tr}\left[{\bf N_{\it i}^{-1}} {\bm\Phi}_i({\bf \Sigma}^{\tilde{f}\tilde{f}}_{kj} +\hat{\tilde{{\bf f}}}_k \hat{\tilde{{\bf f}}}_j^\dagger){\bm \Phi^\dagger_{\it i}}\right]\right\}\\\label{eq:psify} &&\rho^{\tilde{f}y}_i[j]={\rm Re}\left\{\textrm{Tr}\left[{\bf N_{\it i}^{-1}} {\bm\Phi}_i  \hat{\tilde{{\bf f}}}_j{\bf y}^\dagger_i  \right]\right\}\,.
\end{eqnarray}
 One can verify that, by setting ${\bm\Phi}_i$ as an identity matrix, for $i=1,2,\cdots,n_\spara$, such that no instrumental effects are present, this is in complete agreement with the structure of the expression for the mixing matrix in the literature ~\citep{Snoussi:2001bw,2003MNRAS.346.1089D}. 


The problem is evaluating the matrix $\bm \Sigma$ from its inverse as given by Eq.~\ref{eq:Sigma} in order to obtain the covariance matrix ${\bf C}$ from Eq.~\ref{eq:upcl}. Since the matrix $\bm \Sigma$ has the dimensions of $n_\sperp(n_\spara+n_c)\times n_\sperp(n_\spara+n_c)$, which is typically of order $10^6\times10^6$ in a 21-cm survey, such matrix inversion is completely prohibitive computationally. However, as the power spectrum is only determined by the diagonal components of ${\bm \Sigma}$, not by the off-diagonals, by using the same trick presented in the Gibbs sampling techniques~\citep{2004PhRvD..70h3511W,2004ApJ...609....1J} for CMB data analysis, one can verify that the ensemble-averaged solution for the following linear equation has the desired covariance. By solving
\begin{equation}\label{eq:diagRxx}
[{\bf H}^\dagger {\bf N}^{-1} {\bf H}+ {\bf C}^{-1} ]{\bf z} = {\bf H}^\dagger {\bf N}^{-1/2} {\bf z}_1+ {\bf C}^{-1/2}{\bf z}_2 
\end{equation}
where the real vectors ${\bf z}_1$ and ${\bf z}_2$ are of length $n_\sperp n_\spara$ and $n_\sperp(n_c+n_\spara)$, respectively, with elements drawn from a standard normal distribution, one can find that  ${\bf\Sigma}= \left<{\bf zz^\dagger}\right>$ where $\left<\cdot\right>$ represents the average over all solutions for a large number of realizations of ${\bf z}_1$ and ${\bf z}_2$. This algorithm can rapidly provide accurate approximate solutions in massively parallel computers. Therefore one can obtain the diagonal components of ${\bf\widehat{R}_{xx}}$ in Eq.~\ref{eq:upcl} efficiently by $({\bf\widehat{R}_{xx}})_{ii}= \left<|z_i|^2\right> + |\hat{x}_i|^2$ where $i$ runs over all elements.  In the same way as in ${\bf \hat{x}}$, the vector ${\bf z}$ with length $(n_\spara+n_c)n_\sperp$ can also be equally split into $(n_\spara+n_c)$ subvectors, each with length $n_\sperp$, as ${\bf z}=\left(({\bf z}_1^{\tilde{s}})^T,\cdots,({\bf z}_{n_\spara}^{\tilde{s}})^T,({\bf z}_1^{\tilde{f}})^T,\cdots,({\bf z}_{n_c}^{\tilde{f}})^T\right)^T$, in correspondence with contributions from the 21-cm signal, denoted by ${\bf z}_i^{\tilde{s}}$ for $i=1,\cdots,n_\sperp$, and the ICs, denoted by ${\bf z}_i^{\tilde{f}}$ for $i=1,\cdots,n_c$. Note that, in general the noise could be correlated, resulting in non-zero off-diagonal components in the noise covariance matrix, ${\bf N}$.  Solving Eq.~\ref{eq:diagRxx} will require multiplication by a dense $n_{\sperp}n_{\spara}\times n_{\sperp}n_{\spara}$ inverse noise covariance matrix, with a computational scaling of $\mathcal{O}(n^2_{\sperp}n^2_{\spara})$. In the most general case, inverting ${\bf N}$ is a significant computational challenge. However, the inverse noise covariance matrix needs only to be computed once and stored, and then can be used many times to compute Eqs.~\ref{eq:psiff}, \ref{eq:psify} and \ref{eq:diagRxx}.

Using the above mentioned trick, the trace term in Eq.~\ref{eq:psiff} can be computed merely by the ensemble averaged dot products of two vectors as 
\begin{eqnarray} 
&&\textrm{Tr}\left[{\bf N_{\it i}^{-1}}{\bm\Phi}_i({\bf \Sigma}^{\tilde{f}\tilde{f}}_{kj}+ \hat{\tilde{{\bf f}}}_k \hat{\tilde{{\bf f}}}_j^\dagger) {\bm \Phi^\dagger}_i\right]= \nonumber\\
&&\left< ({\bm\Phi}_i{\bf z}^{\tilde{f}}_j)^\dagger({\bf N_{\it i}^{-1}}{\bm\Phi}_i{\bf z}^{\tilde{f}}_k)\right>+ ({\bm\Phi}_i \hat{\tilde{{\bf f}}}_j)^\dagger{\bf N_{\it i}^{-1}}({\bm\Phi}_i \hat{\tilde{{\bf f}}}_k) \, , 
\end{eqnarray}
where the vectors ${\bf z}^{\tilde{f}}_j$ and ${\bf z}^{\tilde{f}}_k$ represent the subvectors of ${\bf z}$ solved by Eq.~\ref{eq:diagRxx}. The above equation applies also to the case of correlated noise.

Moreover, in each EM iteration, we fix the magnitude of each column of ${\bf M}$ to unity and adjust the corresponding power spectra of the ICs accordingly, similar to~\cite{2003MNRAS.346.1089D}, to break the degeneracy and keep the scale-invariant product $\sum_j M_{ij}M_{i'j}\left<\tilde{f}_j[u_k]\tilde{f}^*_j[u_k]\right> =\sum_j M_{ij}M_{i'j}C_f^j[u_k] $  unchanged for an arbitrary pixel $u_k$ at any frequency channels $i$ and $i'$, as seen from Eq.~\ref{eq:modeluv}.

After updating ${\bf M}$ and ${\bf C}$, since the quantities, $\ell^2C^i_\ell$ for the ICs and $k_\spara k_\sperp^2 P_{\rm HI}$ for the HI signal are expected to vary more slowly than  $C^i_\ell$ and $P_{\rm HI}$ themselves, it is more appropriate to perform bin averages after each iteration and update the associated elements in ${\bf C}$, while ensuring the number of observations be smaller than the number of the estimated parameters in ${\bf C}$ and ${\bf M}$. The simulation shows that using band-averaged spectra can significantly increase the convergence speed and obtain much more stable spectra with high accuracy in iterations. According to Eq~\ref{eq:pku}, the bin-averages then yield the following estimates: 
\begin{eqnarray}
C_f^j ({\bf u}) \leftarrow \left(\sum_{|{\bf u}|\in D_q}|{\bf u}|^2\right)^{-1} \sum_{|{\bf u}|\in D_q} |{\bf u}|^2 C_f^j({\bf u})~~~~~\\
 P_{\rm HI} ({\bf u,\eta}) \leftarrow \Bigg(\sum_{{\begin{array}{c}_{|{\bf u}|\in D_q} \\_{\eta\in D_z}\end{array}}}\eta|{\bf u}|^2 \Bigg)^{-1} \sum_{\begin{array}{c}_{|{\bf u}|\in D_q} \\ _{\eta\in D_z}\end{array}} \eta|{\bf u}|^2 P_{\rm HI}({\bf u},\eta)~~~~~
\end{eqnarray}
where $D_q$ is the set of ${\bf u}$ values contributing to bin $q$ and $D_z$ is the set of $\eta$ values contributing to bin $z$. Alternatively, if the 21-cm signal is expected to be highly isotropic, it is also appropriate to perform bin averages by averaging over spherical shells of constant $|k|$ to obtain the spherically averaged HI power spectrum. The bin-averaging schemes mentioned above are especially necessary at the beginning of iterations with a poor initial guess since bin averaging can smooth out any unreasonable values and highly suppress  foreground contamination that is expected to significantly increase in strength towards low $k_\spara$ and would result in a substantial overestimate of HI power spectrum.  One has to keep using the bin averaging process until the derived amplitude of the HI signal in each bin converges to a reasonable range.

In addition, although the bin-widths can be chosen arbitrarily, an appropriate choice can reduce the correlation between the band-power estimates while keeping accurate detection on the structure of the power spectrum. The minimum bin-width can be approximated by the typical correlation length in Fourier space. For an interferometer, one may choose a bin-width of perpendicular modes, $\Delta {\bf u}$, greater than the characteristic width of the Fourier transformed primary beam pattern $A({\bf x})$, and a bin-width of parallel modes, $\Delta \eta$, roughly greater than the inverse of the bandwidth. 

The EM algorithm iterates until there is no significant change in the likelihood. The stopping criterion of the iterative process is here set in terms of the relative change of the bin-averaged HI power spectrum in the last iterations, i.e., $|P_{\rm HI}^{n+1} -P_{\rm HI}^{n} |/|P_{\rm HI}^{n}|< 10^{-3}$ for all bins, and a typical number of iterations is a few hundred.

\subsection{The initial guess}
Since the EM algorithm is a hill-climbing approach, the searching procedure may fail to reach a global maximum and instead converge on a local maximum if the likelihood functions are not convex. In this study, we adapt a commonly used strategy to solve this problem. We try many different initial values varied in reasonable ranges and choose the solution that has the highest converged likelihood value. After a careful investigation, we chose the following algorithm for initialization. For the initial value of ${\bf M}$, we perform PCA analysis~\cite{2008MNRAS.388..247D} to obtain the $n_c$ eigenvectors associated with the largest eigenvalues for the frequency-frequency covariance matrix estimated by averaging over all pixels of data, and use those vectors as the $n_c$ columns of the mixing matrix ${\bf M}$. After the mixing matrix is initialized, the power spectra $C^j_f$ are chosen to be the corresponding diagonal elements in ${\bf \hat{x}} ({\bf \hat{x}})^\dagger$, which is solved using Eq.~\ref{eq:wf} and setting ${\bf C^{-1}}=0$ such that there is no prior information about ${\bf C}$;  Eq.~\ref{eq:wf} reduces to a standard map-making equation. After initializing ${\bf M}$ and $C^j_f$, we assume a flat power spectrum with amplitude comparable to the noise level as prior information about the HI signal. The simulations show that the HI signal reconstruction is quite insensitive to the initial guess for its power spectrum as long as it is not unreasonably large.      

\section{simulations}\label{sect:sim}
We perform simulations to generate dirty sky map data cubes. These data cubes include realistic models for the cosmological 21-cm signal, several diffuse foreground components, and instrumental noise. As this is just the first test of HIEMICA, 
 for simplicity we assume that the primary beam is unity for all frequency channels and assume complete uv-coverage. We leave detailed investigations about more realistic observations to future work.

Because an interferometer only measures temperature fluctuations around the mean and is insensitive to the mean value of brightness temperature, we set the mean of the 3D HI signal and the means of the foreground and noise maps at each frequency channel to zero. 

The 3D dirty sky map is simulated in a box with $64^3$ pixels (i.e., $64$ pixels per side), covering a $30^\circ\times30^\circ$ sky patch and spanning over $780-880$ MHz at intervals of $1.59$ MHz, corresponding to redshifts between 0.82 and 0.61. In comoving coordinates, this box corresponds to about $1341\times 1341\times 606~\text{Mpc}^3$ and the size of each pixel is about $21\times21\times9.5~\text{Mpc}^3$. The code was run using the best-fit cosmological parameters from the Planck measurements~\citep{2014A&A...571A..16P}. 

\subsection{HI Signal}
In cosmology the power spectrum is typically represented in the $({\bf k_\sperp},k_\spara)$ comoving coordinates. If the observed frequency band is small enough (i.e., probing a small range in redshifts) and one uses the flat-sky approximation, there is a linear mapping between these variables:
\begin{equation}\label{eq:mapping}
 {\bf u} =\frac{{\bf k}_{\sperp}{D_{c}(z)}}{2\pi} ; \eta \approx \frac{c\,(1+z)^{2}}{2\pi H_0\nu_{21}E(z)} k_\spara,
\end{equation}
where  $E(z) \equiv \sqrt{\Omega_m(1+z)^{3} + \Omega_\Lambda}$, $\nu_{21}$ is the rest frequency of the 21-cm line, $D_{c}$ is the transverse comoving distance, $z$ is the redshift of the observation, $H_0$ is the Hubble parameter, $c$ is the speed of light,  and $\Omega_m$ and $\Omega_\Lambda$ are the normalized matter and dark energy density, respectively. The angular wavenumber ${\bf k}_\sperp$ and the parallel wavenumber $k_\spara$ are the components of the wavenumber ${\bf k}$ perpendicular and parallel to the line-of-sight direction, respectively. Therefore inserting Eq.~\ref{eq:mapping} into \ref{eq:pku}, we obtain the relation of the power spectrum defined under the different coordinates:
\begin{equation}\label{eq:pkk}
P_{\rm HI}({\bf k_\sperp},k_\spara) =\frac{c\,(1+z)^{2}}{H_0\nu_{21}E(z)} P_{\rm HI}({\bf u},\eta).
\end{equation}

The 21-cm brightness temperature and the corresponding 3D power spectrum can be written as:

\begin{eqnarray}
P_{\rm HI}(k) & = & \left( \bar{T}_{\rm HI}(z)  \right)^2 \, b^2(k,z)\, D^2(z) P_{\text{cdm}}(k)  \\
 \bar{T}_{21}(z)  & \simeq & 0.084  \, \mathrm{mK}  
\frac{ (1+z)^2}{E(z)} 
 \dfrac{\Omega_B}{0.044}  \,  \frac{f_{\rm HI}(z)}{0.01} 
\end{eqnarray}
where $b(k,z)$ is the bias parameter and $P_{\text{cdm}}$ is the cold dark matter power spectrum at the present day. $\Omega_B$ is the baryon density fraction. $D(z)$ is the growth factor for dark matter perturbations defined such that $D(0)=1$.  For the purpose of this paper, we assume $b=1$ over redshift and scale. For simplicity, we also neglect the effects of redshift-space distortions caused by the peculiar velocities of HI clouds and galaxies in the HI power spectrum since the ICA-based approach is insensitive to the detailed shape of the power spectrum. The HI mass fraction is expected to increase with redshift and we assume a linear dependence: $f_{\rm HI}(z)=0.008(1+z)$. The matter power spectrum $P_{\text{cdm}}$ with the baryon acoustic oscillations (BAO) can be parametrized according to the simple empirical formula:
\begin{eqnarray} \label{eq:pk21}
&&\frac{P_{\text{cdm}}(k_\sperp,k_\spara)}{P_{\rm ref}(k_\sperp,k_\spara)} = 1+\\ 
&& A\, k \exp \left( -(k/\tau)^\alpha\right)
\sin\left( 2\pi\sqrt{\frac{k^2_\sperp}{k^2_{\rm BAO\sperp}} + 
\frac{k^2_\spara}{k^2_{\rm BAO\spara}}}\;\right)~~
\end{eqnarray}

where $k=\sqrt{k^2_\sperp +k^2_{\spara}}$, the parameters $A$, $\alpha$ and $\tau$ are adjusted according to the formula presented in~\cite{1998ApJ...496..605E}. $P_{\rm ref}(k)$ is the smooth ``no-wiggles'' power spectrum at $z=0$, which can be computed from the fitting formula given by~\cite{1998ApJ...496..605E}.

The parameters $k_{\rm BAO\sperp}$ and $k_{\rm BAO\spara}$ are the sinusodial scales in the radial and transverse directions in k-space. We choose the following values for these parameters used in this paper: $A=1.0$, $\tau=0.1~h\mathrm{Mpc^{-1}}$, $\alpha=1.4$ and $k_{\rm BAO\sperp}=k_{\rm BAO\spara}=0.060~h\mathrm{Mpc^{-1}}$.

\subsection{Foregrounds}
\begin{figure}[h]
  \centering
  \includegraphics[width=3.3in]{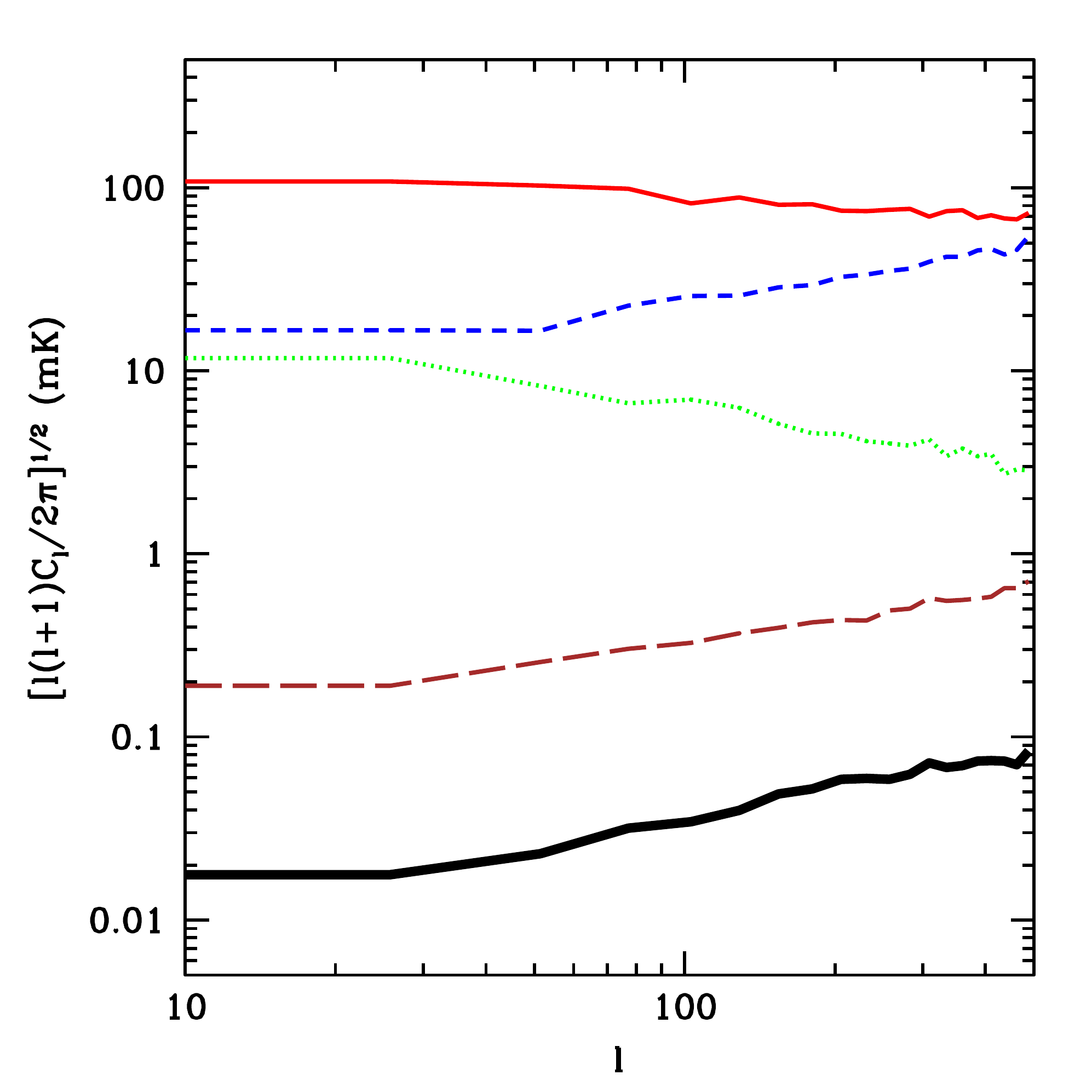}
  \caption {The angular power spectra derived from the simulated maps of the 21-cm signal (black-thick) and foregrounds at $z= 0.71$ ($\nu=830$ MHz), with a bandwidth of $\Delta \nu = 1.59$ MHz. The foreground components include Galactic synchrotron radiation (red-solid), extragalactic point sources (blue-dashed), and Galactic free-free (green-dotted) and extragalactic free-free (brown long-dashed). }
\label{fig:cl}
\end{figure}

We model the foregrounds as isotropic random Gaussian fields described by angular power spectra $C_\ell(\nu,\nu')$ based on~\cite{2005ApJ...625..575S}. In this paper,  we assume that bright resolved point sources have been removed accurately and consider the dominant four diffuse components: Galactic synchrotron emission, Galactic and extra-galactic free-free emission, and extragalactic radio point sources. The angular power spectrum of each source takes the generic form:
\begin{equation}\label{eq:fcl}
C_f(\ell,\nu,\nu') = A\left( \frac{1000}{\ell} \right)^\beta \left(\frac{\nu^2_f}{\nu\nu'}\right)^{2\alpha}\exp\left(-\frac{\ln^2(\nu/\nu')}{2\xi^2}\right) \,
\end{equation}
where  $\nu_f$ is the reference frequency with $\nu_f=130~$MHz.
We list the parameters of the foreground models used in this paper in Tab.~\ref{tab:cl}. Based on such models, Fig.~\ref{fig:cl} shows the angular power spectrum of each foreground contribution for one realization. The corresponding maps are shown in Fig.~\ref{fig:diffF}.

\begin{table}
\begin{center}
\begin{tabular}{l|c|c|c|c|c}
\hline
&$A ({\rm mK}^2)$ & $\beta$ & $\alpha$  & $\xi$ & $\Delta \alpha$\\
\hline
extragalactic point sources   &  57.0  & 1.1 &  2.07 & 1.0 &0.2\\
extragalactic free-free        &  0.014  & 1.0 &  2.10 & 35 &0.03 \\
Galactic synchrotron          &  700  & 2.4 &  2.80 & 4.0 &0.15 \\
Galactic free-free          &  0.088  & 3.0 &  2.15 & 35  &0.03  \\
\hline
\end{tabular}
\end{center}
\caption{Foreground model parameters for angular power spectrum $C_\ell(\nu,\nu')$ used in Eq.~\ref{eq:fcl}.}
\label{tab:cl}
\end{table}

Following~\cite{2006MNRAS.370.1125D}, the frequency index $\alpha$ of real-world foregrounds varies slightly across the sky. For example, in the case of synchrotron emission, the direction-dependent spectral index reflects variations of the relativistic electrons density and Galactic magnetic field. In this paper, for each foreground component, we assume the indices in different directions to be Gaussian distributed with a mean of $\alpha$ and an rms of $\Delta \alpha$ as shown in Tab.~\ref{tab:cl}, consistent with model parameters chosen in~\cite{2012MNRAS.419.3491L}.

\subsection{Noise}

For each frequency channel, we assume Gaussian noise in the $uv$-plane with zero mean and that the noise maps at two different frequencies are uncorrelated. We make noise realizations in the $uv$-plane for a given noise power spectrum for each frequency. The thermal noise of the measurement in units of brightness temperature is given by~\citep{2006ApJ...653..815M,2012A&A...540A.129A}
\begin{eqnarray}\label{eq:Nuf}
P_{\rm N}({\bf u},\nu) &=& \left<|\tilde{n}({\bf u},\nu)|^2\right> =\left(\frac{\lambda^2}{A_e}\right)^2 \frac{ T^2_{sys}}{\Delta\nu \Delta t n_b}\\
&\simeq& 1.9\times10^{-7}\,{\rm mK}^2 (1+z_\nu)^4\left(\frac{T_{sys}}{50\,{\rm K}}\right)^2\nonumber\\
&\times&\left(\frac{10\text{m}}{A_e}\right)\left(\frac{10}{n_b}\right)\left(\frac{1\,{\rm MHz}}{\Delta\nu}\right)\left(\frac{30\,{\rm days}}{\Delta t}\right)\nonumber
\end{eqnarray}
where $z_\nu$ is the corresponding redshift of the frequency $\nu$, $T_{sys}$ is the system temperature,  $n_b$ is the total number of redundant baselines within that $u_k$ pixel with integration time $\Delta t$ for each baseline,  $\Delta \nu$ is the width of the frequency channel, and $A_e$ is the effective area of each individual antenna in the array.

In order to keep the results as general as possible, we do not consider a specific configuration of the array;  the main goal in this paper is not to discuss instrumental effects in 21-cm power spectrum measurements. Instead, we consider two simple scenarios in our simulation to illustrate the effectiveness of the presented foreground-cleaning approach.  One scenario assumes a currently achievable noise level of $\sqrt{P_{\rm N}}\simeq1\times10^{-3}$ mK for each pixel of the $uv$-plane, and the other one assumes a noise level of about $2.2\times10^{-4}$ mK for future measurements. The former has an averaged signal-to-noise ratio over the whole data cube about 1 while the latter is about 5.

\section{Applications to simulated data}\label{sect:app}
In this Section we show the results obtained by applying the HIEMICA algorithm to the simulated data. There are two main findings, namely the map reconstruction and the power spectrum recovery. As the fluctuations of the cosmological 21-cm signal are expected to be isotropic, we focus on the recovery of the averaged 3D power spectrum in spherical shells of constant $|k|$. However, redshift-space distortions will slightly break spherical symmetry, and we will investigate the consequences in a future work. The estimated power spectra are obtained by averaging over the results from 10 independent realizations of the simulated data cubes and the associated statistical errors are obtained from their dispersions. Each data cube is for a fixed sky patch that includes the same simulated foregrounds combined with independent realizations of the instrumental noise and the HI signal.


\begin{figure*}[htbp]
\centering

\mbox{
\subfigure[Galactic synchrotron emission]{
   \includegraphics[width=2.3in] {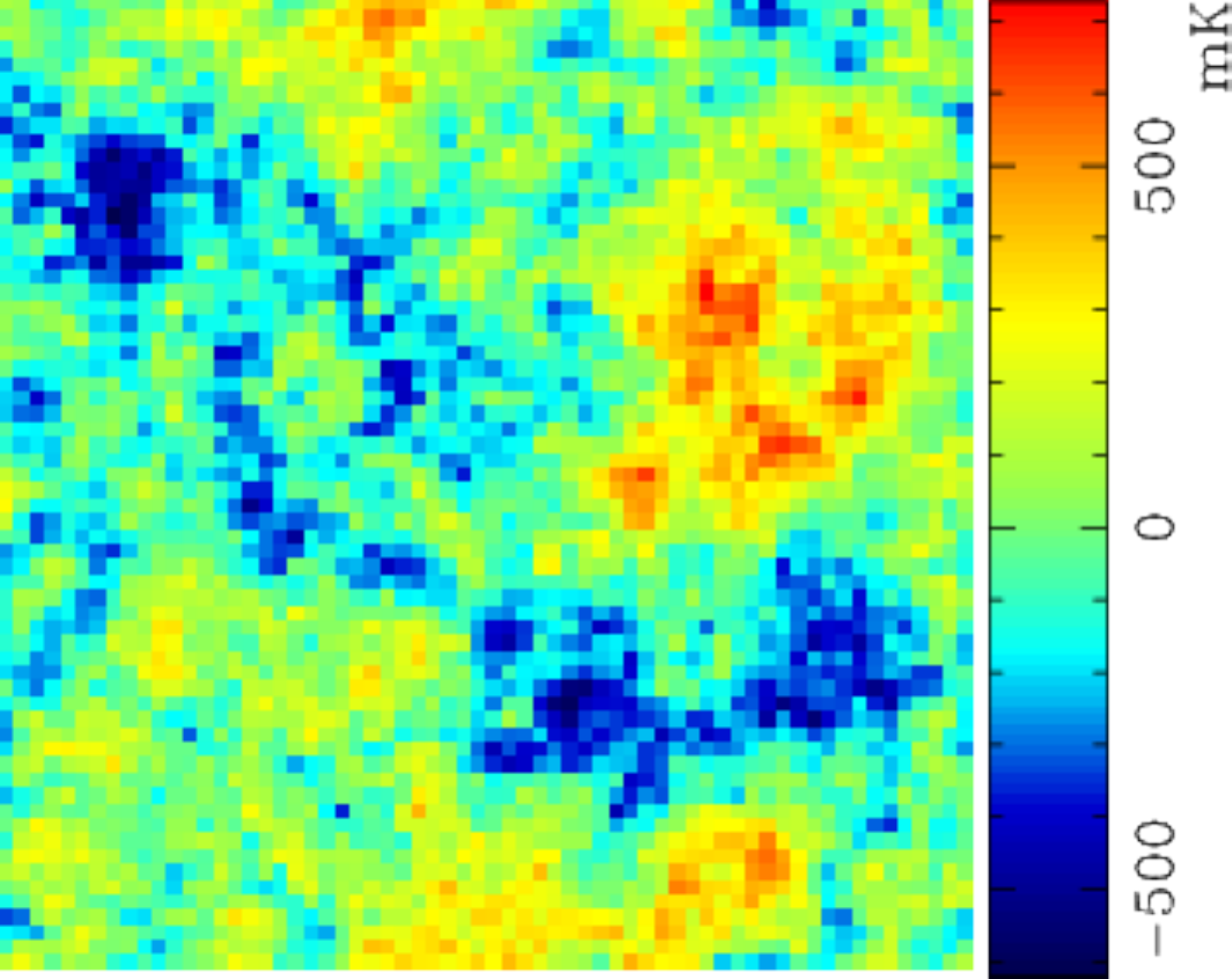}
   \label{fig:Gs} }

\subfigure[Extragalactic point sources]{
   \includegraphics[width=2.3in] {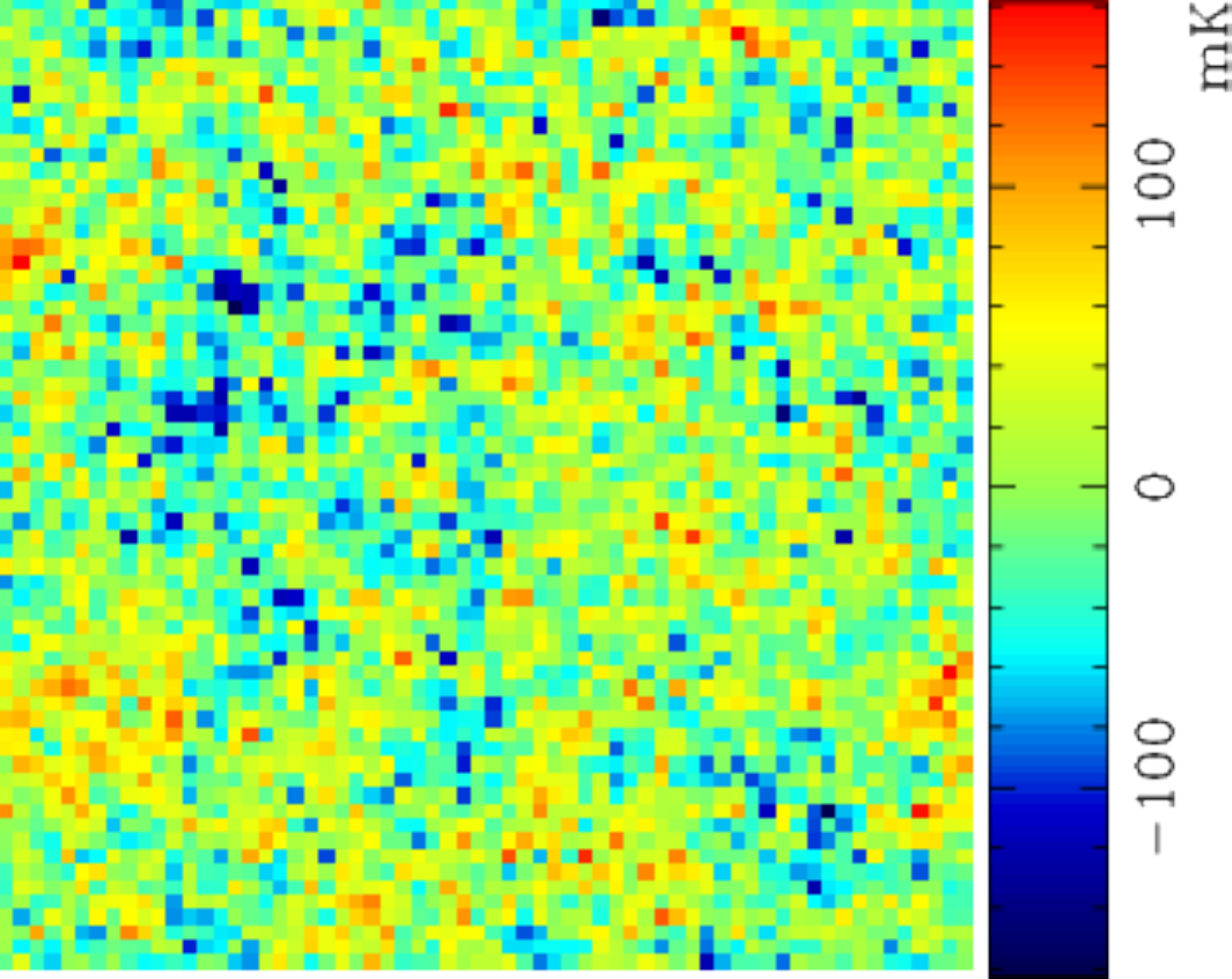}
   \label{fig:Ep} }
}
\mbox{
\subfigure[Galactic free-free emission]{
   \includegraphics[width=2.3in] {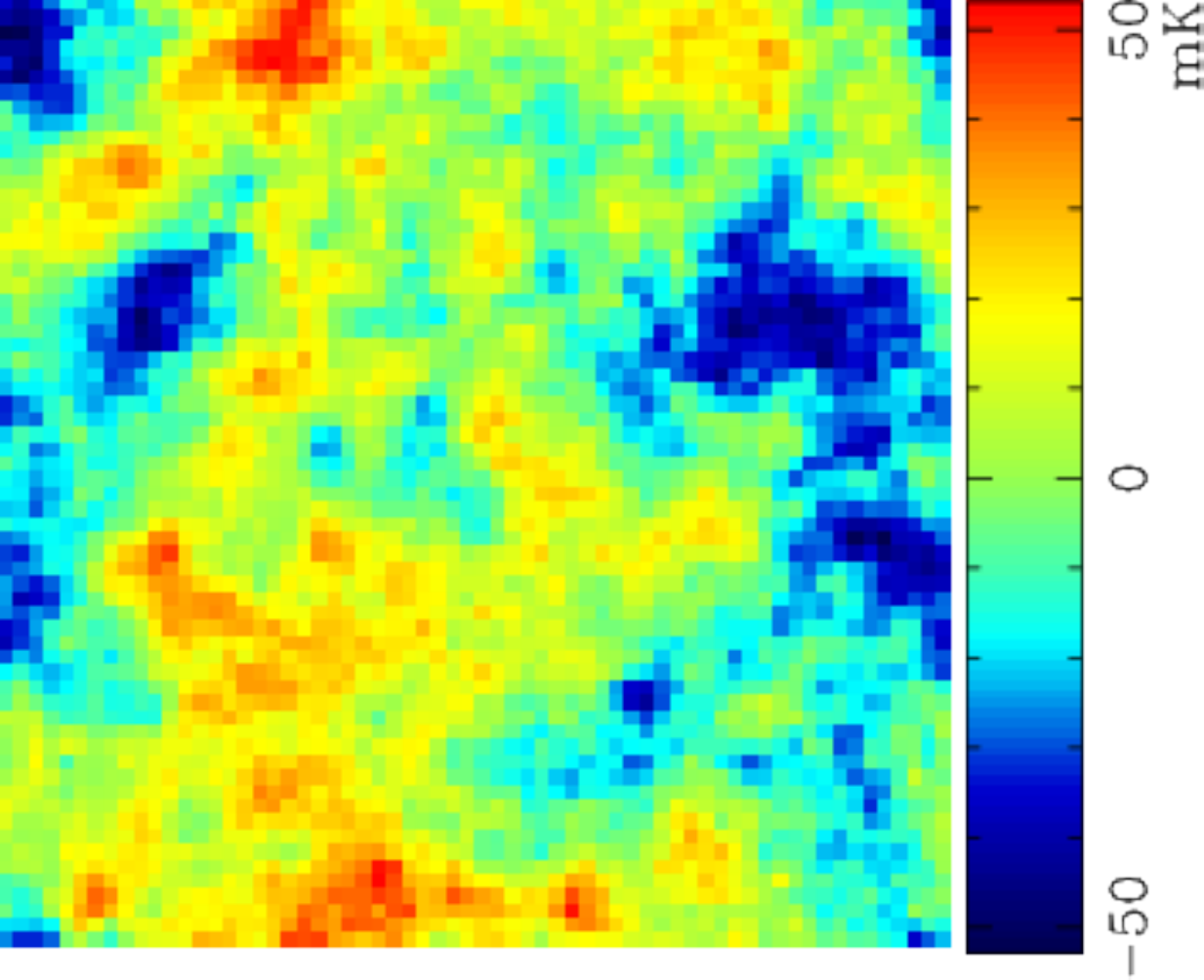}
   \label{fig:Gf} }

\subfigure[Extragalactic free-free emission]{
   \includegraphics[width=2.3in] {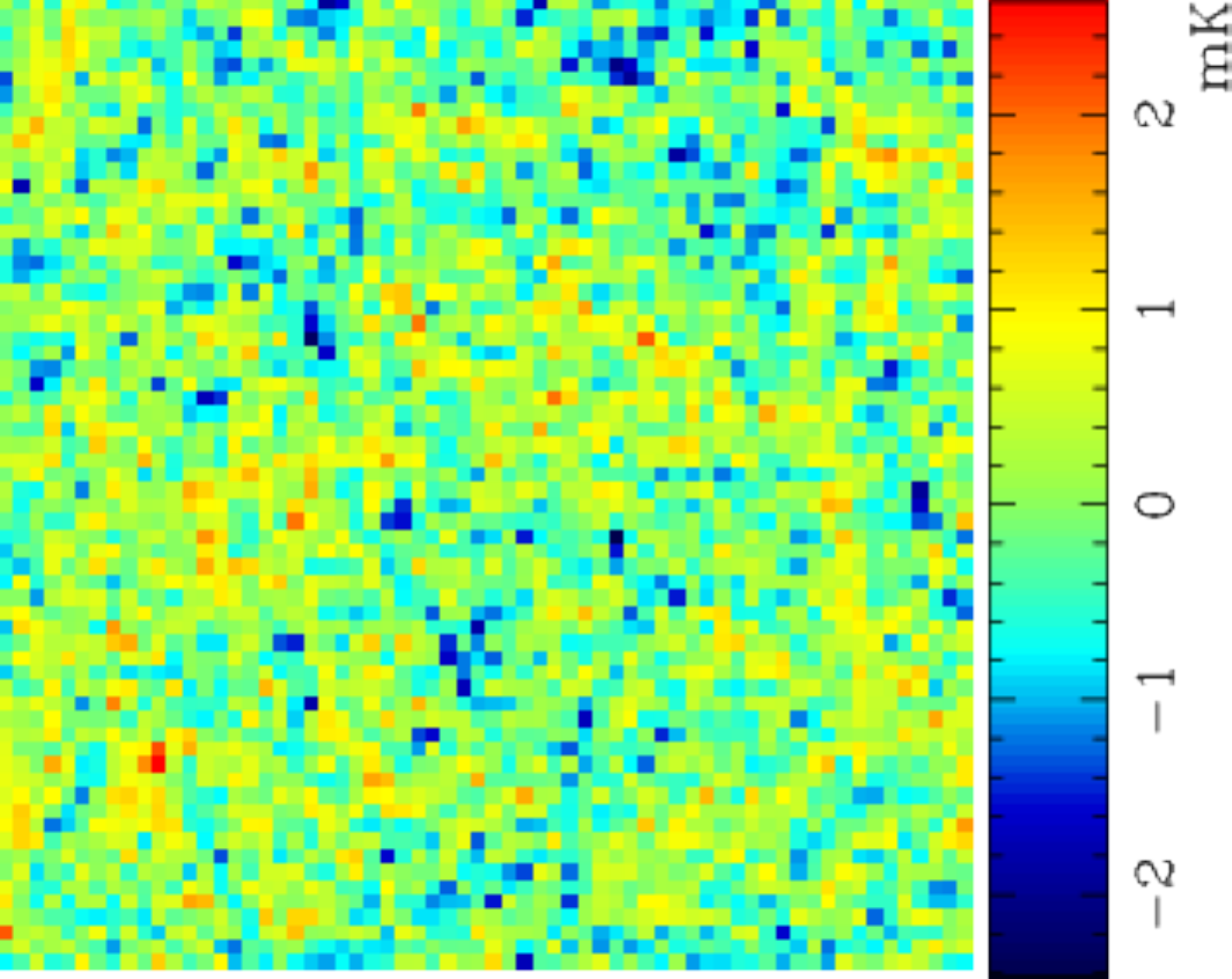}
   \label{fig:Ef} }
}
\caption{The simulated diffuse foreground components for a $30^\circ\times 30^\circ$ patch of sky at $\nu=830$ MHz. We have subtracted the mean value (i.e. DC mode) to reflect an interferometric observation.}
\label{fig:diffF}
\end{figure*}

\begin{figure*}[htbp]
\centering
\mbox{
\subfigure[The first foreground component]{
\includegraphics[width=2.3in] {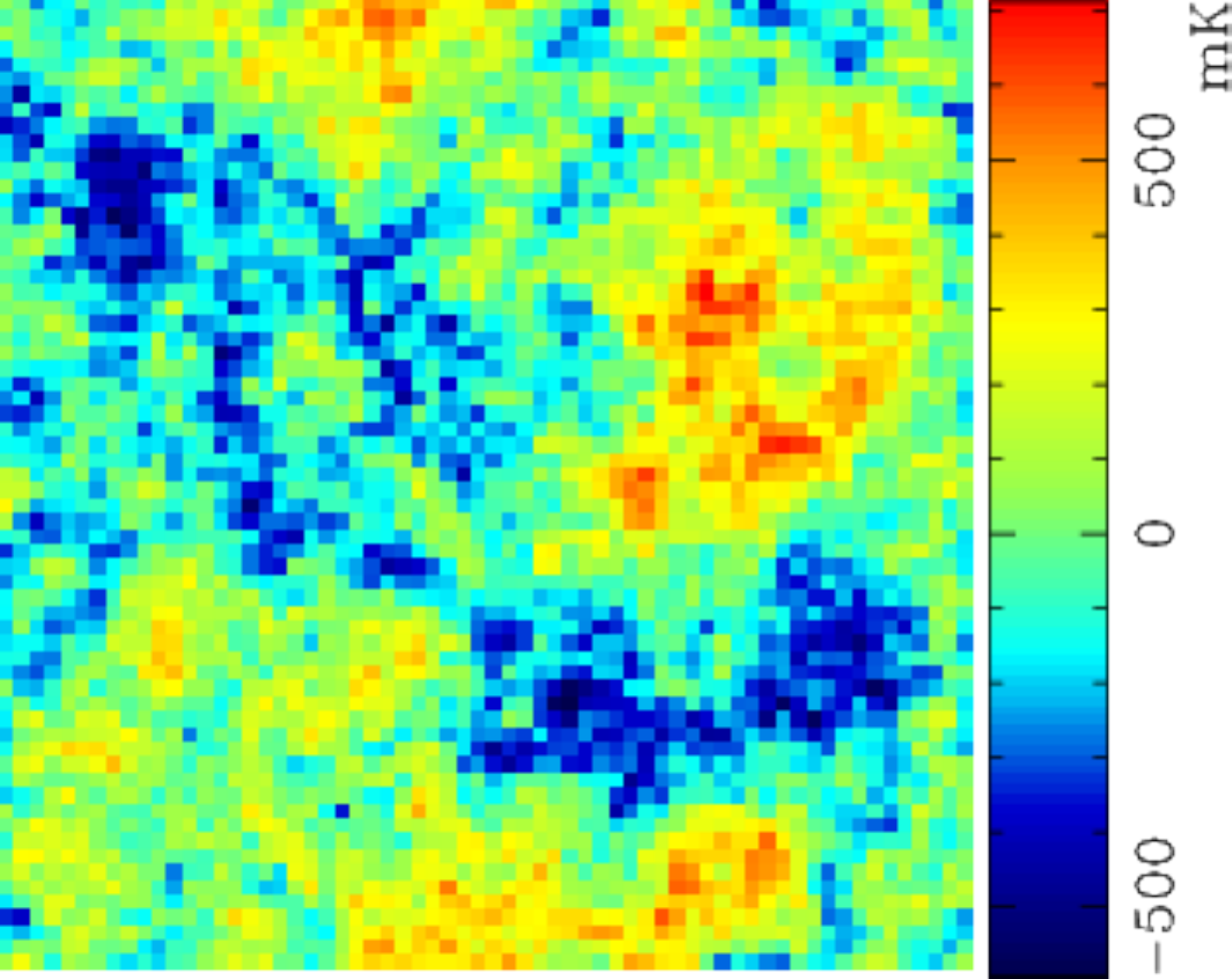}
 \label{fig:F1} } 

 \subfigure[The second foreground component]{
 \includegraphics[width=2.3in] {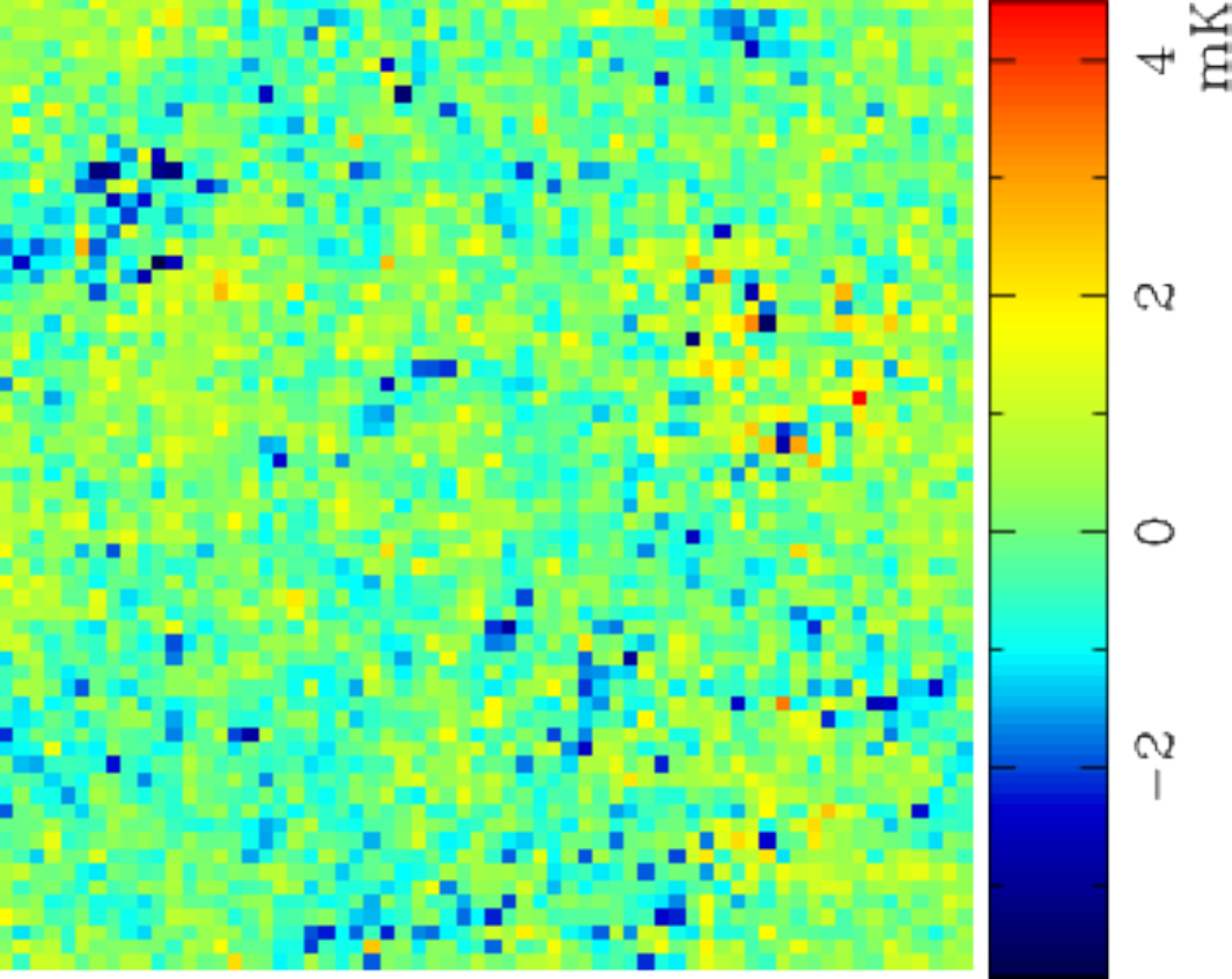}
   \label{fig:F2} }  }

\mbox{
 \subfigure[The third foreground component]{
 \includegraphics[width=2.3in] {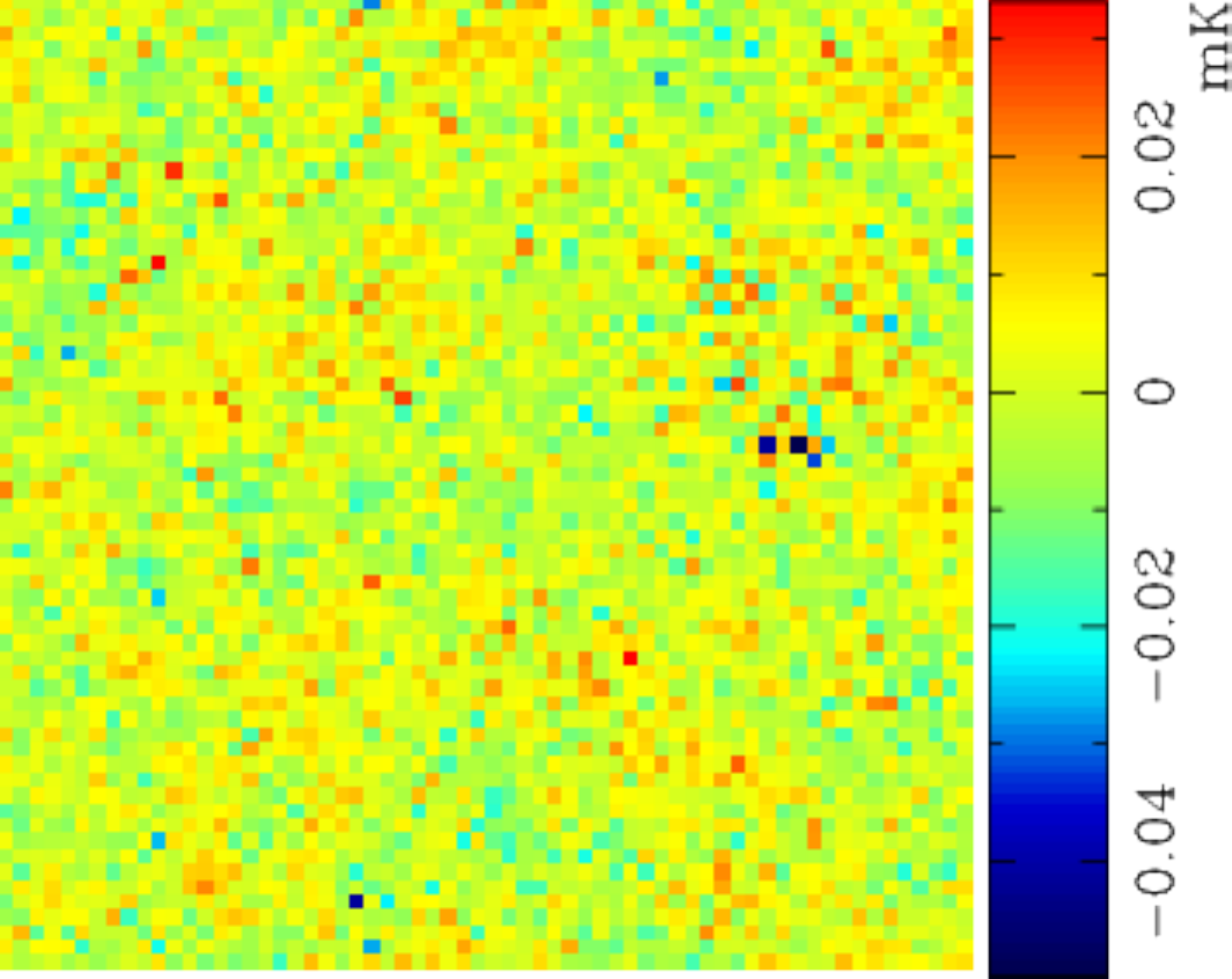}
   \label{fig:F3} } 
 \subfigure[The fourth foreground component]{
 \includegraphics[width=2.3in] {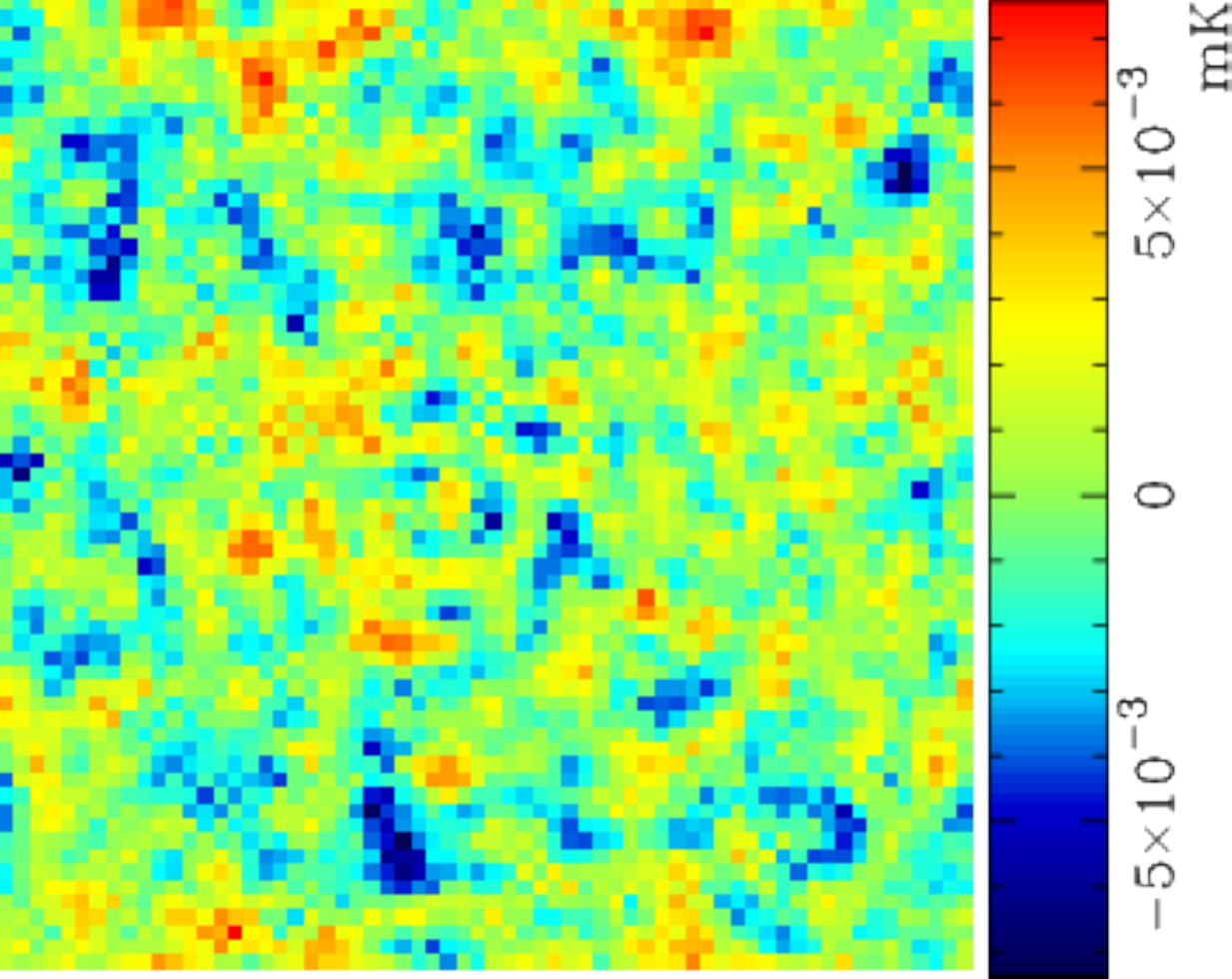}
   \label{fig:F4}} }

\caption{Wiener-filtered maps of independent components (ICs) reconstructed by applying HIEMICA to the simulated data cube with S/N=5 for the same sky patch and frequency as in Fig.~\ref{fig:diffF}. Notice that we do not expect the recovered maps to explicitly correspond to the input maps since these ICs are assumed to be mutually independent while the true physical components may have correlations.}
\label{fig:wfF}
\end{figure*}

\begin{figure}[htpb!]
 \centering
 \includegraphics[width=3.5in]{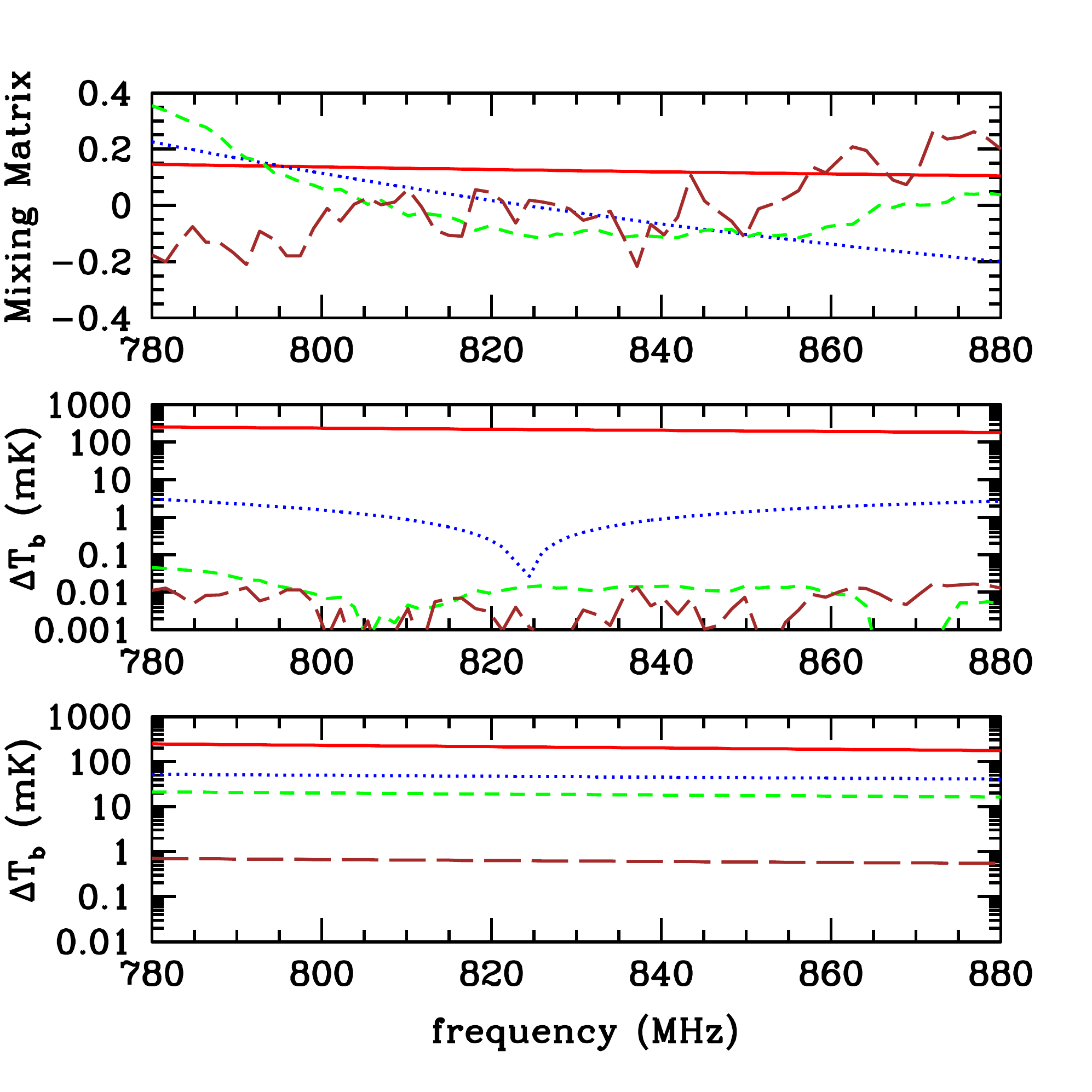}
 \caption{{\bf Top panel}: HIEMICA-derived coefficients of each column of the mixing matrix, representing the frequency dependence of the ICs, where the red-solid line is for the first component, blue-dotted for the second, green-dashed for the third and brown long-dashed for the fourth. When applying HIEMICA to simulated data cubes, here we have assumed $n_c$=4 and S/N=5. {\bf Middle panel}: the rms brightness temperatures of ICs calculated from the derived mixing matrix and their angular power spectra. {\bf Bottom panel}: the rms brightness temperatures of the input  physical foreground components including Galactic synchrotron radiation (red-solid), extragalactic point sources (blue-dashed), and Galactic free-free (green-dotted) and extragalactic free-free (brown long-dashed). As mentioned in Fig.~\ref{fig:wfF}, the temperature fluctuations of the ICs are not necessarily the same as the physical components. }
\label{fig:mix}
\end{figure}

\begin{figure*}[htbp!]
\centering
\mbox{ \subfigure[Input total foregrounds]{
   \includegraphics[width=2.2in] {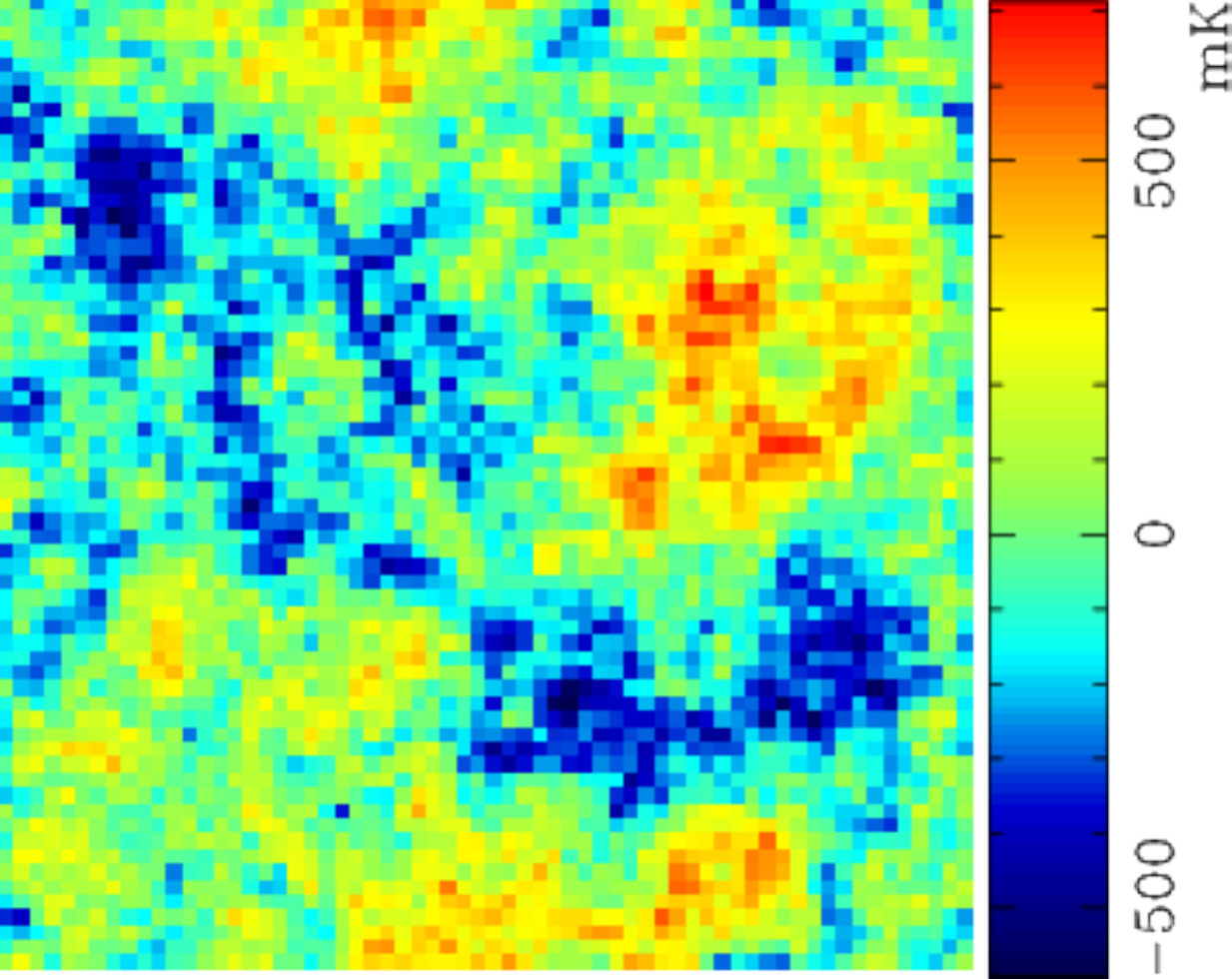}
   \label{fig:F_compare} } 

\subfigure[Recovered total foregrounds]{
\includegraphics[width=2.2in] {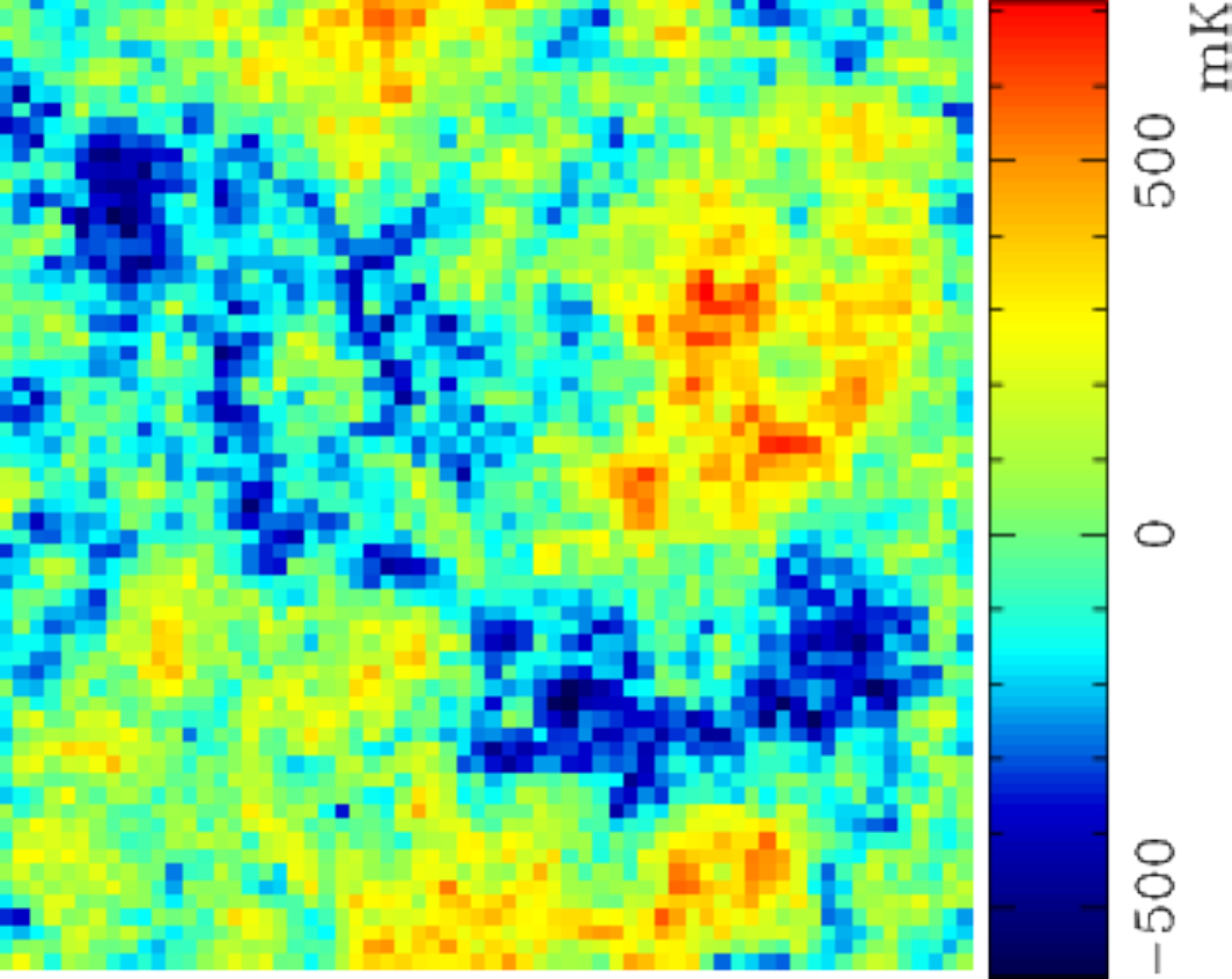}
 \label{fig:F1} }
}

\mbox{
\subfigure[Input 21-cm signal]{
   \includegraphics[width=2.2in] {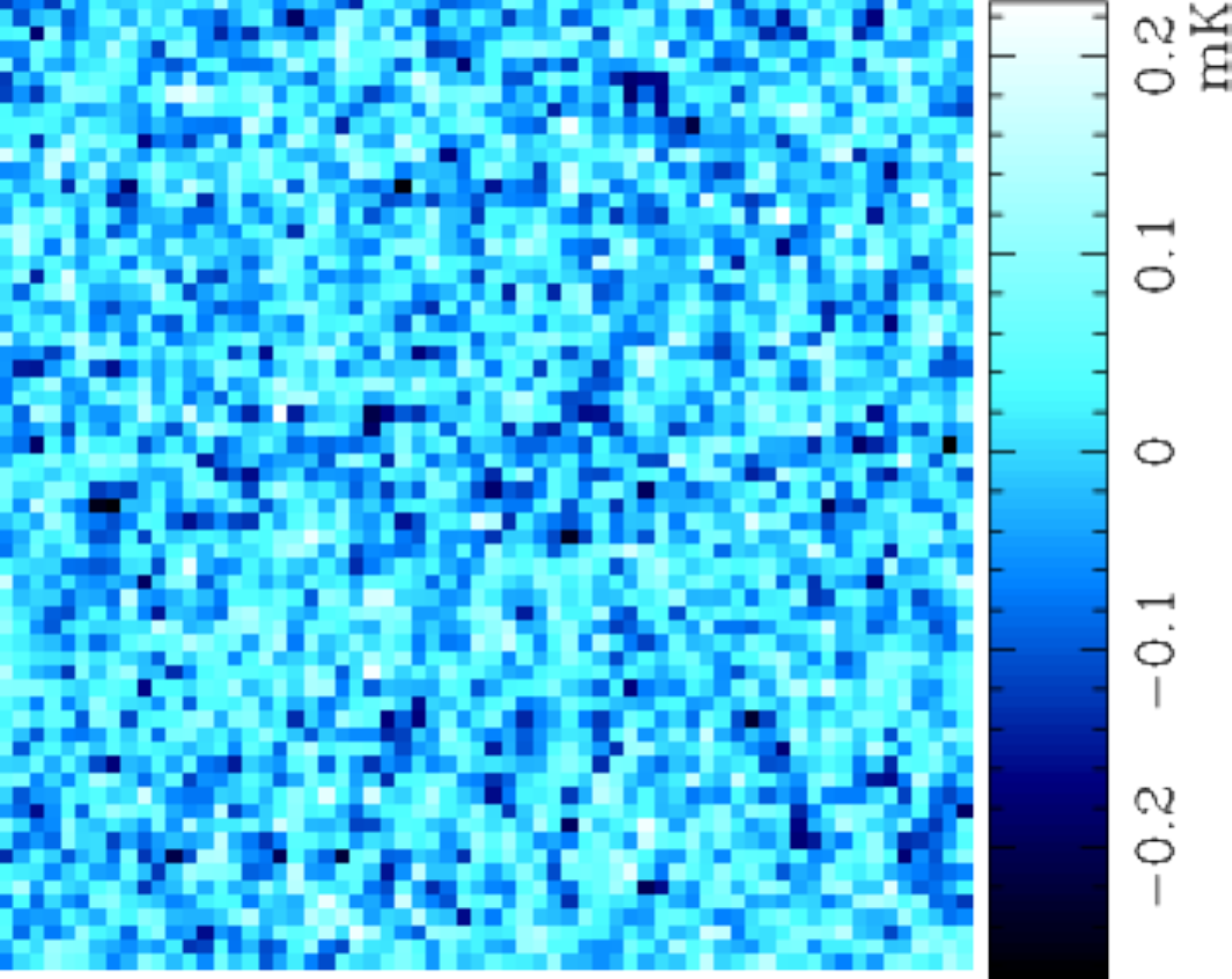}
   \label{fig:21-cm}}

 \subfigure[Recovered 21-cm signal (S/N=1)] {
 \includegraphics[width=2.2in] {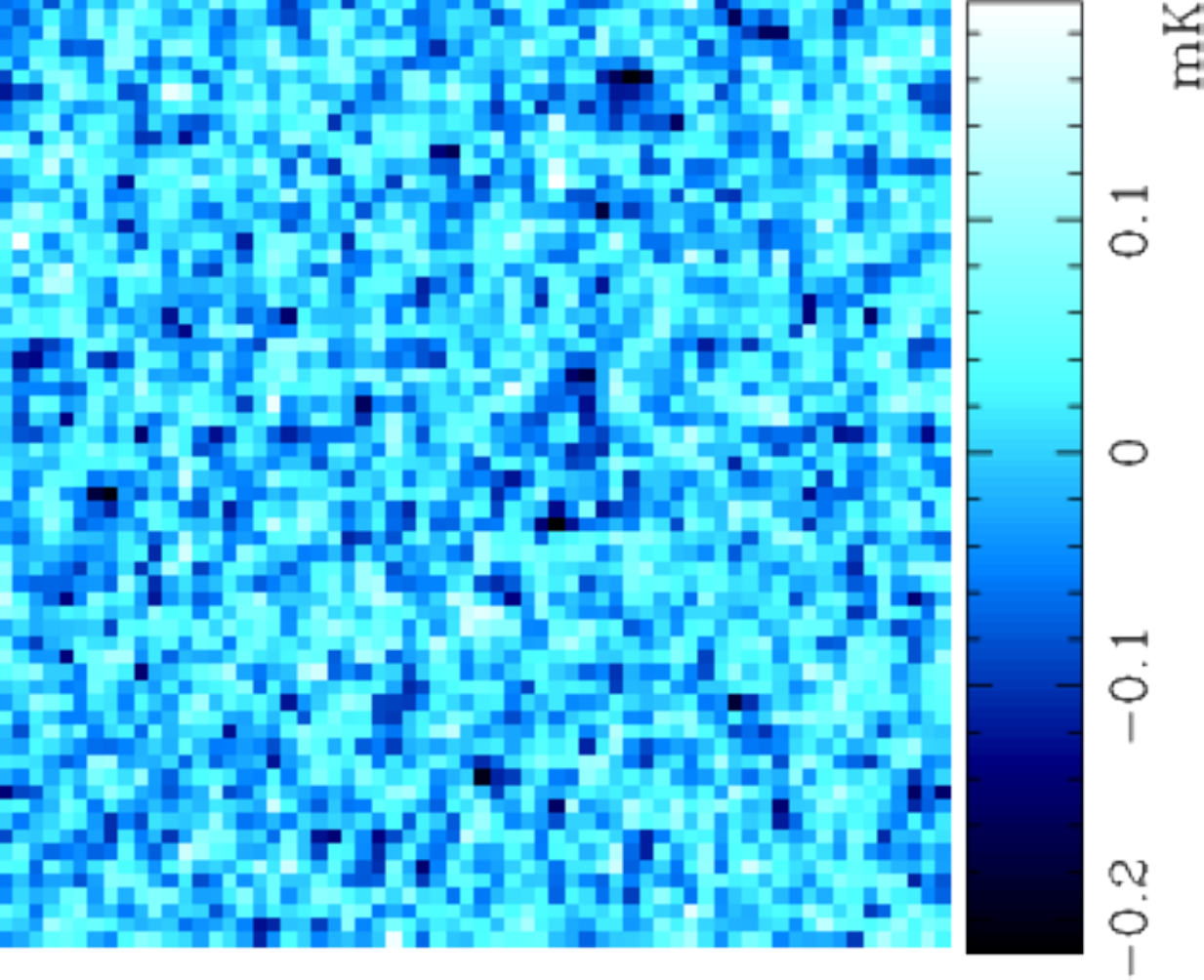}
  \label{fig:r21sn1}}

 \subfigure[Recovered 21-cm signal (S/N=5)] {
 \includegraphics[width=2.2in] {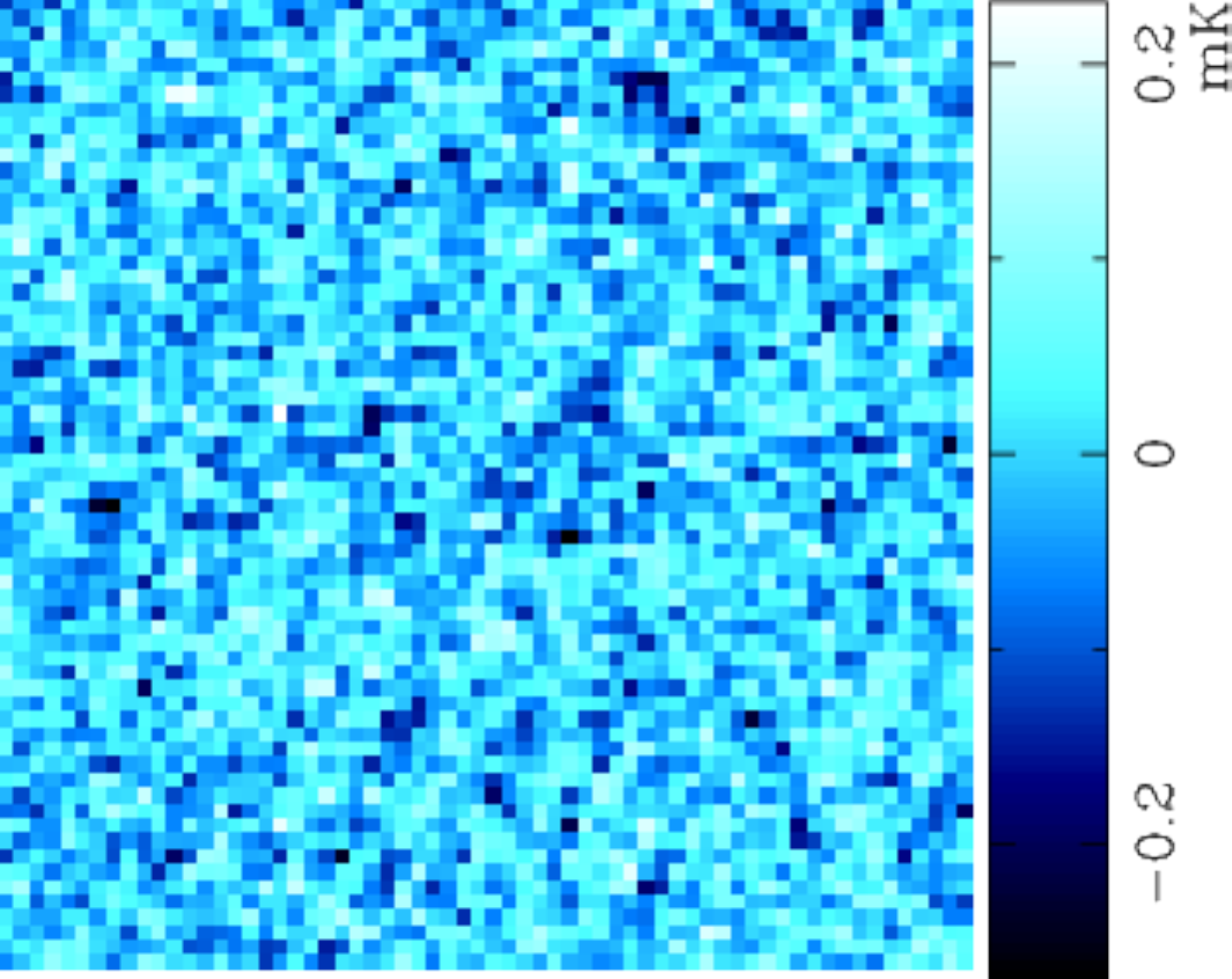}
  \label{fig:r21sn5}}
}

\caption{Same as Fig.~\ref{fig:wfF}, but for the total foregrounds and the 21-cm signal to clearly illustrate that a successful separation was achieved by the HIEMICA-cleaning process. There is an underestimate of the 21-cm signal in the case of S/N=1, which is caused by noise contamination as a Wiener-filtered map $\propto \rm{S/(S+N)}$. An almost perfect recovery occurs over all scales for S/N=5.}
\label{fig:FS-compare}

\end{figure*}

\subsection{The Independent Components}
Fig.~\ref{fig:wfF} shows the four Wiener-filtered ICs at $\nu=830$ MHz found by HIEMICA. The Wiener-filtered components are calculated by Eq.~\ref{eq:wf}, and, the corresponding map of the $i$-th component at the frequency bin $\nu$ is obtained by multiplying such component by the $(\nu,i)$-th entry of the mixing matrix, $M_{\nu,i}$. As seen from Figs.~\ref{fig:diffF}, \ref{fig:wfF} and \ref{fig:mix}, we find that only the first component is compatible with the input Galactic synchrotron contribution, while the others have no corresponding physical foregrounds and have unphysical spectral and spatial behaviours. These unphysical behaviours are due essentially to the fact that the realistic astrophysical components have significant cross-correlations while our independent component analysis assumes mutual independence between the source signals. Thus each separated component is just a mixture of all the source signals, but the sum of these ICs is essentially same as the total astrophysical signal, as seen in Figs.~\ref{fig:F_compare} and ~\ref{fig:F1}. Meanwhile, the Wiener-filtered maps of the 21-cm signal (bottom panel of Fig.~\ref{fig:FS-compare}) show the encouraging result that, the 21-cm signal, even in the case of S/N=1, is nearly perfectly recovered across the map by HIEMICA.

There is a slight difference between the low and high S/N cases (Figs. \ref{fig:r21sn1} and \ref{fig:r21sn5}) at small scales, but no apparent difference at large scales is found. The recovered amplitudes of fluctuations for S/N=1 are smaller than those for S/N=5 by several tens of percent, which is essentially due to noise suppression in Wiener-filtered maps that is $\propto$ S/(S+N), where S/N tends to be low at high-$k$ modes. However, the large-scale fluctuations of the 21-cm signal in both cases are much larger than the noise amplitudes, resulting in almost identical large-scale fluctuations.       

\subsection{Determination of the Number of ICs}
When analyzing data with HIEMICA, we have to choose the number of Independent foreground Components (ICs) to be estimated. Our statistical framework for independent component analysis can provide a rigorous determination of this number by using an empirical statistical approach, the Likelihood Ratio Test (LRT). The LRT provides an objective criterion for selecting between possible models by using the ratio of the likelihoods on the data. Although in practice evaluating likelihood functions is a hard task, we can instead apply HIEMICA several times with successively increasing $N_c$ until the recovered power spectra are essentially unchanged when more components are added into the analysis.  

The LRT statistic is given by $LRT = 2 \ln {\cal L}_s - 2\ln {\cal L}_c$, in which $\ln {\cal L}_s$ and  $\ln {\cal L}_c$ denote the logs of the likelihood functions for a relatively simpler model $s$ (so-called null hypothesis) with fewer parameters and a more complex model $c$ (alternative hypothesis), respectively. Asymptotically, the LRT statistic follows a $\chi^2$ distribution with degrees of freedom equal to the difference in the number of free parameters between the two models. In our case, in analysis of the data with $64$ frequency channels and 30 $\ell$-bins, adding an additional foreground component would increase the degrees of freedom by $64+30 = 94$.

The comparison of the fitting model between the models for $n_c=$ 1 and 2 (i.e., labelled as $1\rightarrow2$ ICs in Tab.~\ref{tab:LRT}) and $n_c=$ 2 and 3 (as $2\rightarrow3$ ICs) shows that, from the values of LRT, $\ln L$ increases significantly as the number of ICs increased up to 3. The significance values ($p\simeq 0$) for S/N of 1 and 5 indicate that only one or two foreground components would result in an unsuccessful fit and the data strongly prefer three foreground components. Moreover, especially in the case of high S/N, there is no statistically significant difference when using more components, given the larger $p$ values ($p > 0.01$). Here all the values of LRT and $p$ are the median over 10 independent simulations. Therefore, it is recommended to use three components in the reconstructed foreground model for our specific simulation parameters in Tab.~\ref{tab:cl}. It is worth noticing that, as mentioned by~\cite{2014arXiv1409.8667A}, the optimal number of ICs in fact strongly depends on the spectral smoothness of true foregrounds, characterized by a frequency correlation length $\xi$ as defined in Eq.~\ref{eq:fcl}. In fact, a longer coherence length implies a smoother frequency spectrum for a physical foreground component. Consequently, a smaller value of $\xi$ may require more ICs to successfully model such physical components and remove them accordingly.


\subsection{3D Power Spectrum Results}
\begin{figure*}[!htbp]
\centering
\mbox{
 \subfigure[S/N=1]{
   \includegraphics[width=3.4in] {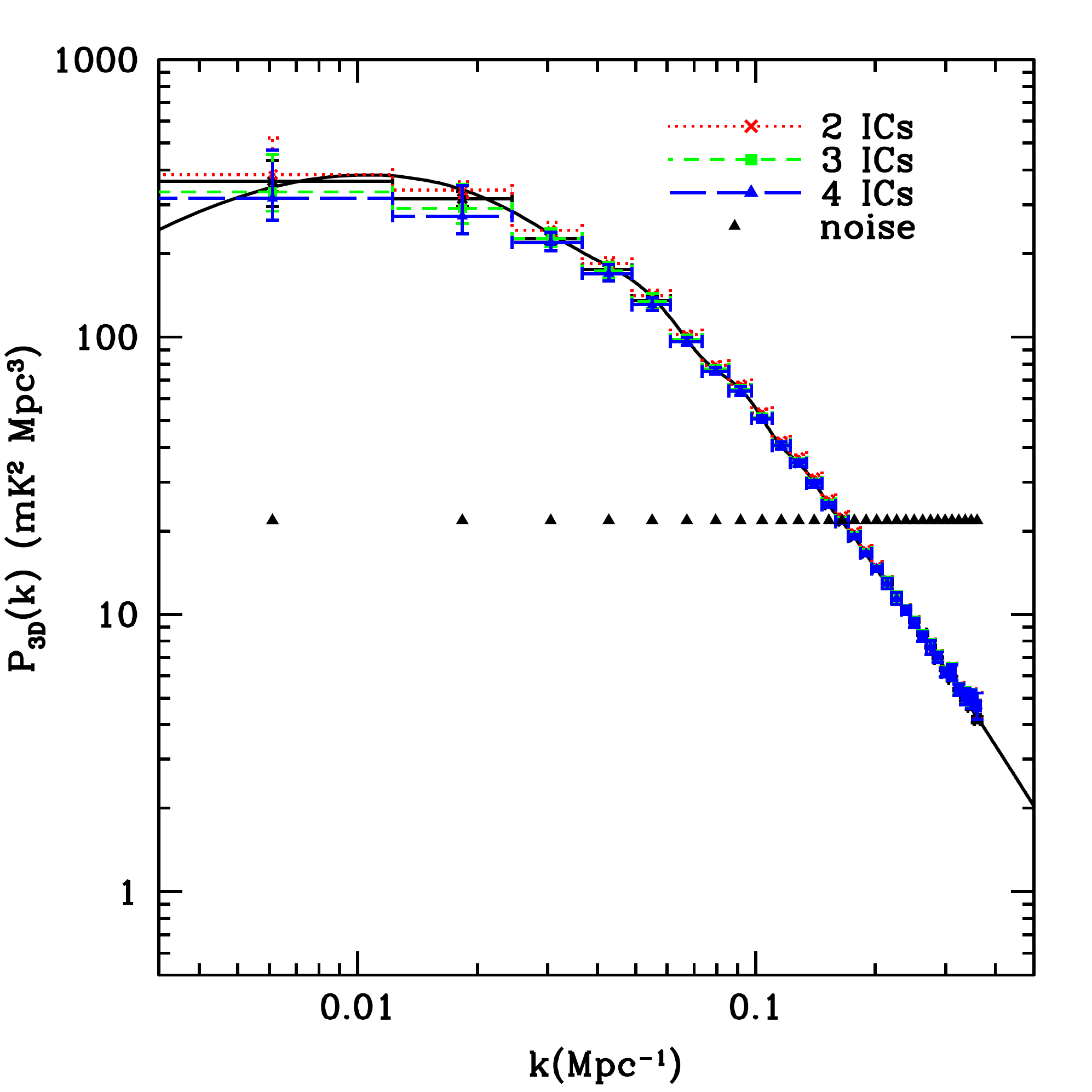}
   \label{fig:compk1} }

\subfigure[S/N=5]{
   \includegraphics[width=3.4in] {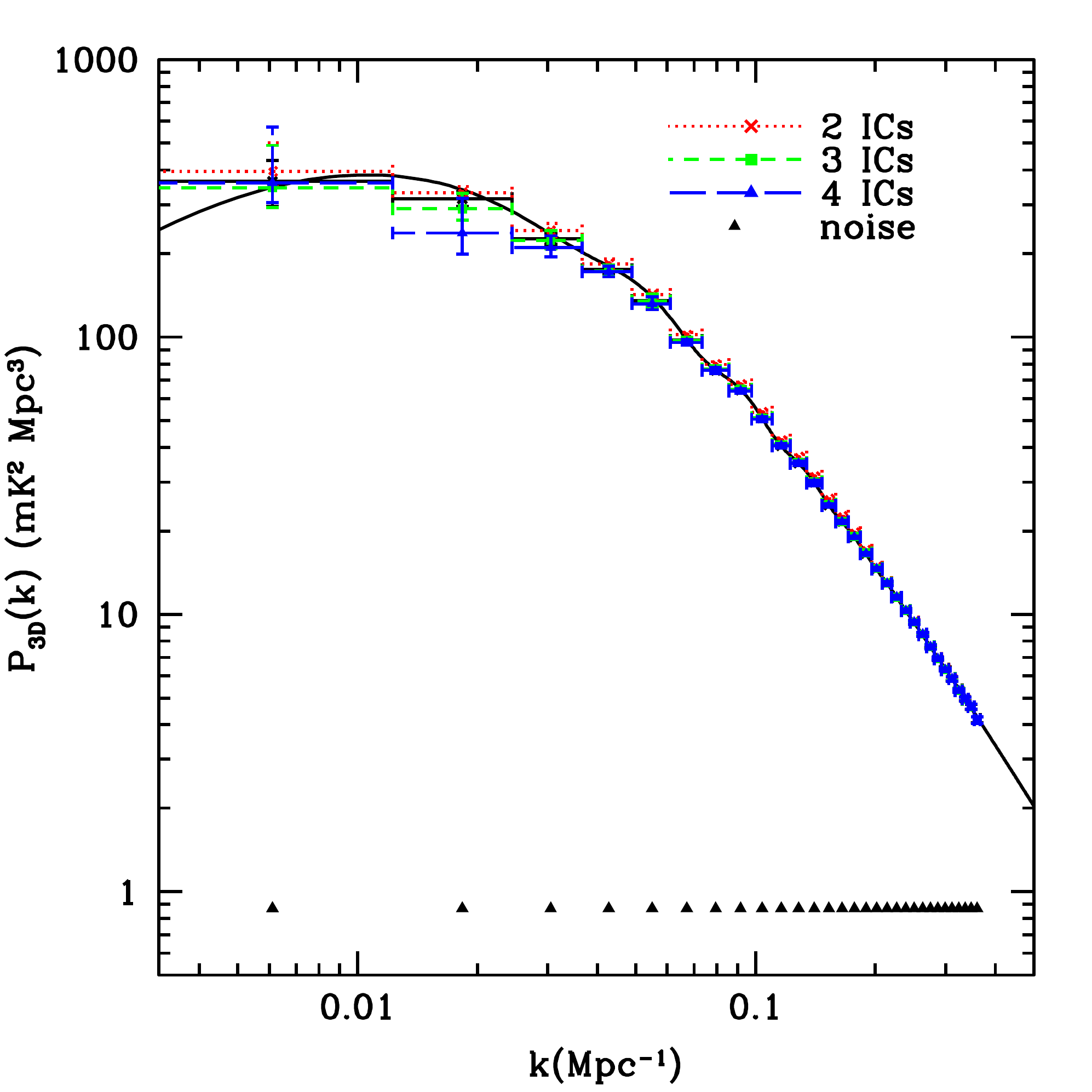}
   \label{fig:compk5}}
}

\mbox{
 \subfigure[S/N=1]{
   \includegraphics[width=3.4in] {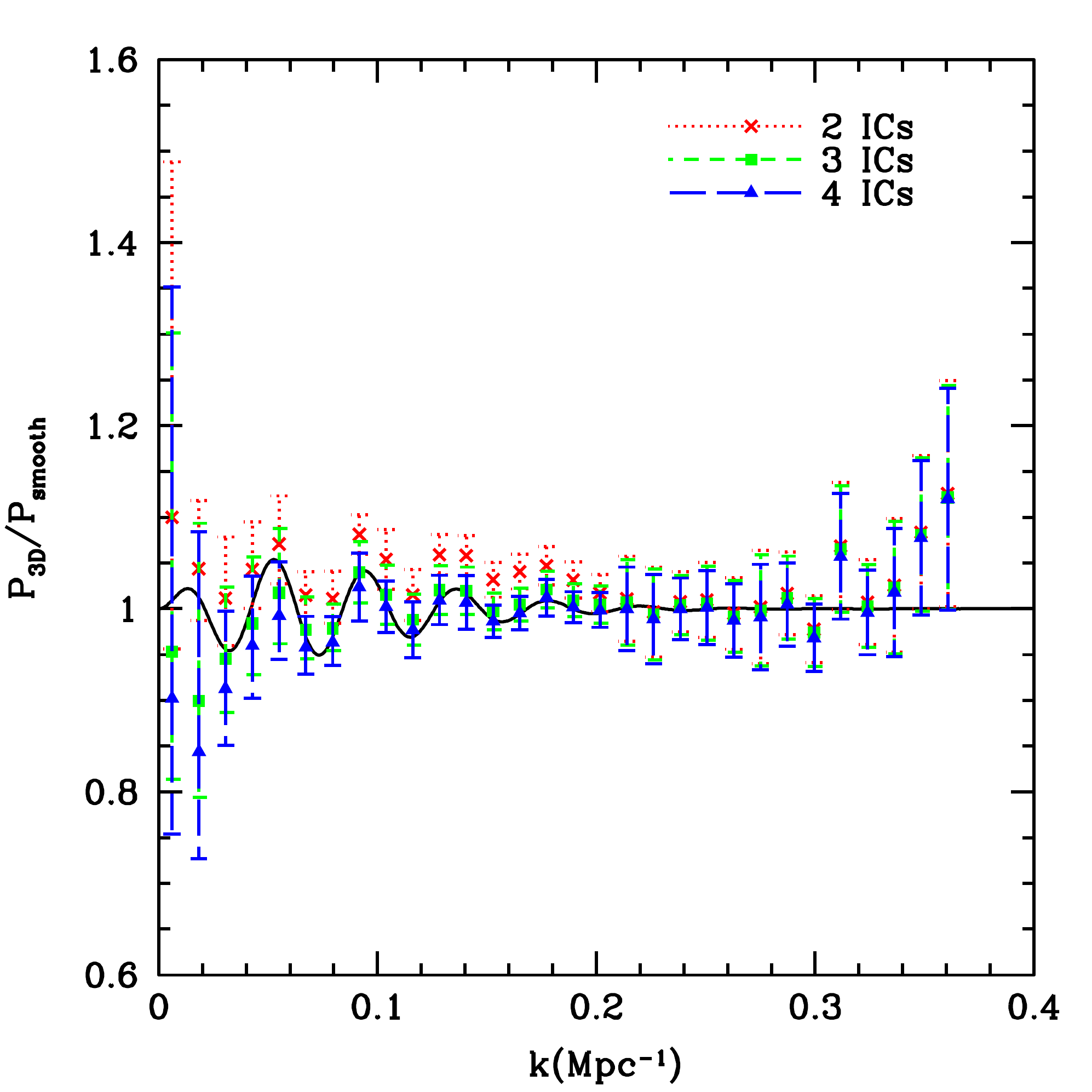}
   \label{fig:compkr1} }

\subfigure[S/N=5]{
   \includegraphics[width=3.4in] {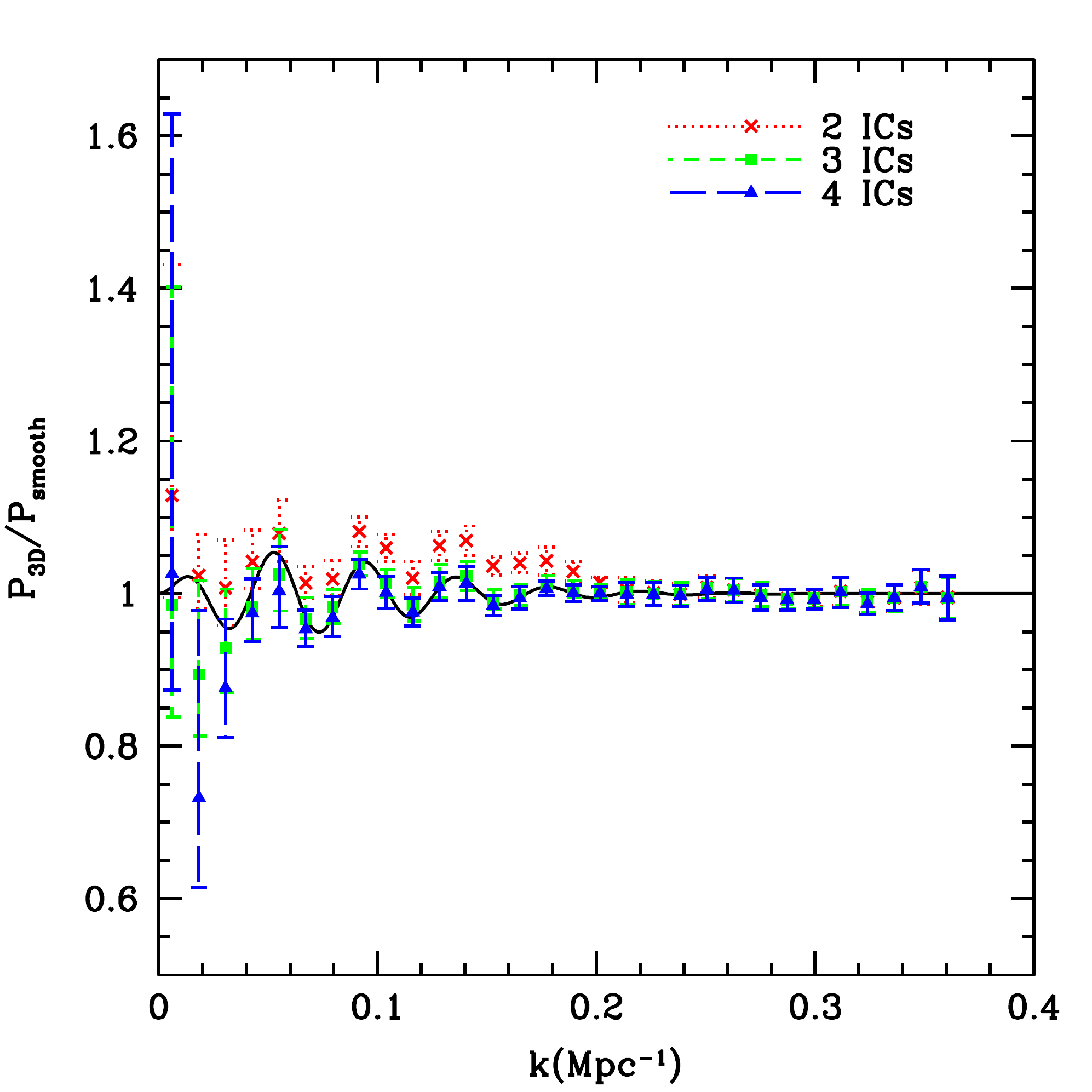}
   \label{fig:compkr5}}
}

\caption{Dependence of the 21-cm signal recovery on the number of independent components, for the data with S/N=1 (left) and 5 (right). {\bf Upper panels}: We show the spherically averaged three-dimensional power spectra of the simulated 21-cm signal(black), noise (black-dotted), reconstructed 21-cm signal for the HIEMICA algorithm run with the assumption of $n_c=2$ (red-dotted), 3 (green-dashed) and 4 (blue long-dashed), respectively. Vertical bars indicate the 1-$\sigma$ errors estimated from 10 realizations and horizontal bars the bin-width of $\Delta k =0.0129$. The statistical uncertainties are typically smaller than the symbol sizes and mostly invisible. {\bf Lower panels}: Same as upper panels, but showing the HI power spectrum divided by a smoothed spectrum without baryon acoustic oscillations to single out those oscillations.               }
\label{fig:compk}
\end{figure*}

Fig.~\ref{fig:compk} shows the rapid convergence of the resulting HI power spectrum when successively increasing the number of ICs.  It can be seen that the recovered power spectrum for IC number of 2 is slightly different from those for IC number of 3 and 4, but the 3 and 4 components result in almost the same power spectrum, which is consistent with the varying trend in the LRT. It can also be confirmed by comparing the reconstructed Wiener-filtered ICs in Fig~\ref{fig:wfF}. As the amplitude of fluctuations of the second component (see Fig.~\ref{fig:F2}) is indeed about 20 times larger than the input 21-cm signal (see Fig.~\ref{fig:21-cm}), the removal of the second component is therefore necessary.  Furthermore, since the averaged amplitude of fluctuations of the third component (see Figs.~\ref{fig:F3} and~\ref{fig:mix}) is about 10 times smaller than the 21-cm signal, the effects of such component in signal recovery are expected to be less important but can still lead to notable improvements at large scales as the corresponding 3D Fourier modes are not significantly smaller than those of the HI signal. Moreover, the amplitude of fluctuations of the fourth component (see Fig.~\ref{fig:F4}) is about two times smaller than the third one, leading to rather small effects in signal reconstruction.

The comparison between the estimated signal power spectra for S/N=1 and 5 in Fig.~\ref{fig:compk}, we note that, for $k\lesssim0.2$ Mpc$^{-1}$ over which the averaged S/N$>1$ for both cases, the impact of the noises at these scales tends to be negligibly small since the recovered power spectra follow the true spectrum very closely and there are no apparent differences between the two cases. On the other hand, the recovered power spectrum for S/N=1 begins gradually to overestimate the true one at $k>0.2$ Mpc$^{-1}$ by up to $12\%$, but for S/N=5 the recovered spectrum is almost same as the true one with only $2\%$ deviation. The additional power in the former case is primarily because the large noise level causes a slight underestimate of the true foreground contributions. The ``excess power'' is barely significant; the simulation truth is still contained in the 1-$\sigma$ error bars.

In Figs. \ref{fig:compkr1} and \ref{fig:compkr5}, we zoom in on the region with BAO features to illustrate clearly the dependence of the power spectrum estimation on the number of ICs as well as on the noise amplitude. We can see that too few components, such as $n_c=2$ in both cases of S/N=1 and 5, results in an overestimate on the power spectrum at scales of $0.1\lesssim k\lesssim 0.2$ Mpc$^{-1}$, since those two components can not completely describe the spectral and spatial properties of the foreground. As the number of ICs increases to 3 and 4, the resulting power spectra decrease and both rapidly converge to the true values.  As the S/N drops down by a factor of 5, the measurement errors become larger since such errors can be approximately estimated by $\sigma(k) \propto \sqrt{P_{\rm HI}(k)+ P_{\rm N}(k)}$. More importantly, one can see that, if $n_c=3$ or $4$, the recovered HI spectrum is well within the 2-$\sigma$ error bars of the estimates.

Notice that we have applied a correction to convert the prediction intervals on the underlying power spectrum into error bars that approximate Bayesian power spectrum inference in the limit of low noise and small numbers of modes. The resulting upper and lower error bars are asymmetric. The detailed analytical estimations of such correction are presented in the Appendix. As a consequence, compared with the uncorrected errors, the upper error bars increased by about $60\%$, $40\%$ and $20\%$ respectively in the lowest three $k$-bins and the associated lower error bars decreased by about $40\%$, $15\%$, $10\%$. The differences are less than a few percent in the higher $k$-bins due to the large number of modes contained within those bins.

\begin{table*}[h]
\centering
\renewcommand{\arraystretch}{1.5}
\caption{Comparison of models by the Likelihood Ratio Test (LRT) for the number of independent foreground components.} \label{tab:LRT} 
\begin{tabular}{|c|c|l|l|l|l|}
\hline
Model &$\Delta$ $dof$ & LRT (S/N=1) & LRT (S/N=5) &$p$-value (S/N=1)& $p$-value (S/N=5) \\
\hline
$1\rightarrow 2$ ICs&94 &17558 &20872&0\footnotemark[$\dagger$]&0\footnotemark[$\dagger$] \\
$2\rightarrow3$ ICs &94 &196   &280  &$<10^{-6}$\footnotemark[$\dagger$]&$<10^{-6}$\footnotemark[$\dagger$] \\
$3\rightarrow4$ ICs &94 &120   &125   &0.036&$0.018$\\

$4\rightarrow5$ ICs &94 &96   &108   &0.42&$0.15$\\
\hline
\end{tabular}
\footnotetext[$\dagger$]{$p<0.01$, typically used to assess the significance of  LRT.}
\end{table*}

\begin{figure*}[h]
\centering
\mbox{
 \subfigure[S/N=1]{
   \includegraphics[width=3.4in] {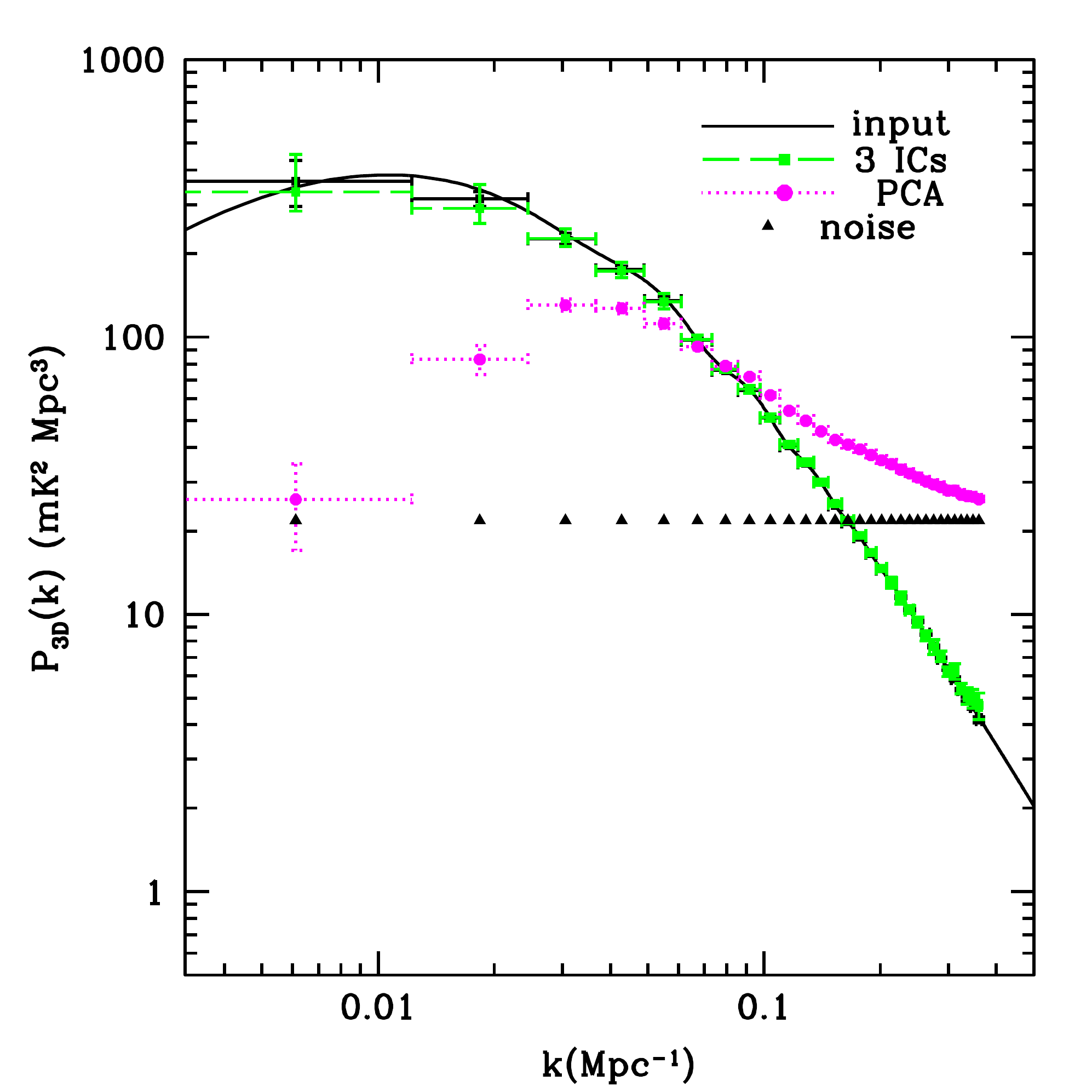}
   \label{fig:pk1} }

\subfigure[S/N=5]{
   \includegraphics[width=3.4in] {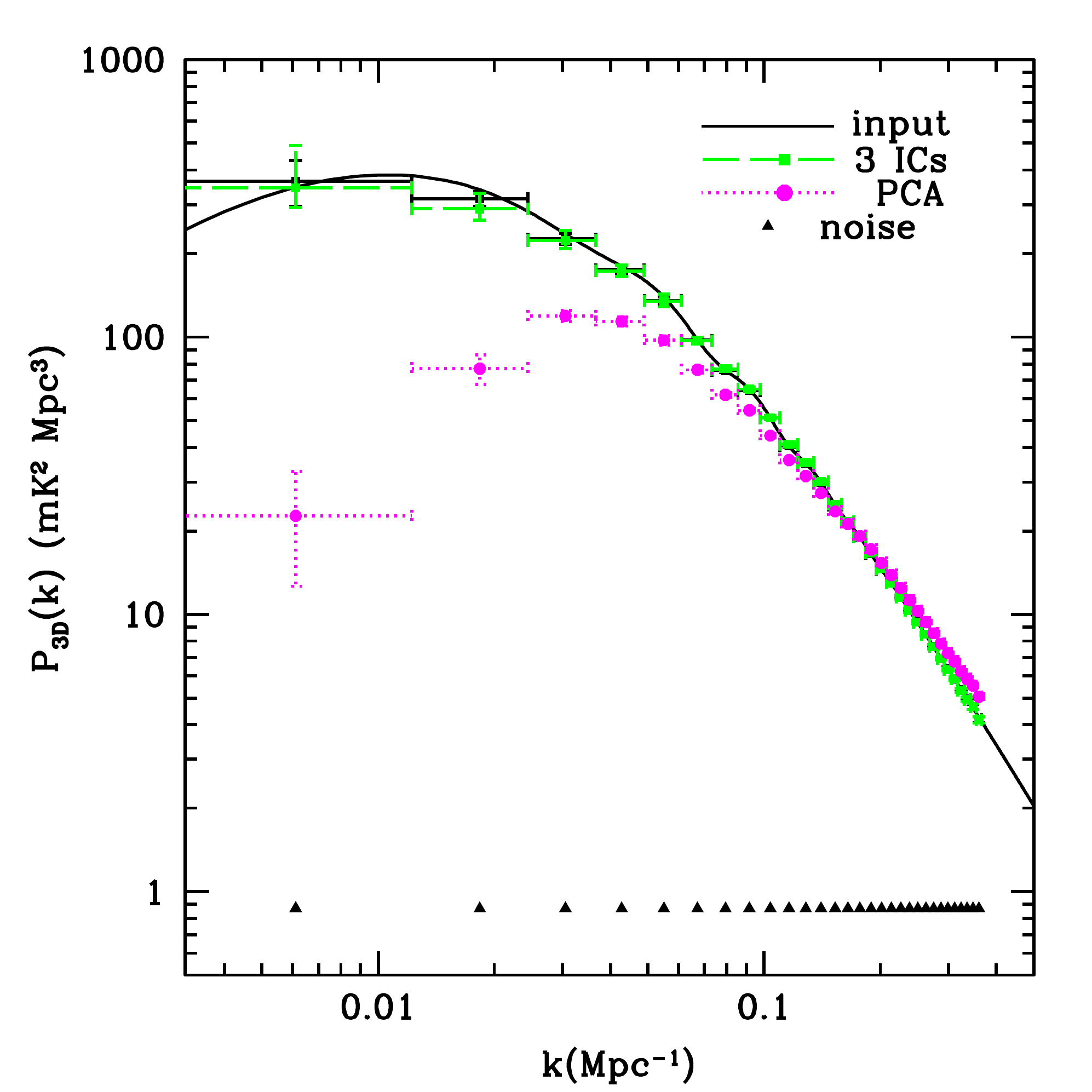}
   \label{fig:pk5}}
}

\caption{Same as Fig.~\ref{fig:compk}, but for comparison of 21-cm power spectrum recovery based on the PCA technique (magenta-dotted) and the HIEMICA approach (gren long-dashed), for the data with S/N=1 (left) and S/N=5 (right). The spherically averaged three-dimensional power spectra of the simulated 21-cm signal(black-solid), noise (black-dotted), HIEMICA-derived (assumed $n_c$=3) 21-cm signal and PCA-derived residuals from projecting out the first three dominant eigen-components based on the frequency-frequency covariance of data.}
\label{fig:pk}
\end{figure*}

We can also see that the recovered cosmological signal decreases approximately monotonically with increasing number of ICs, especially for the low-$k$ modes ($k\lesssim0.1$ Mpc$^{-1}$). This result may be interpreted as follows: (1) since each recovered foreground component is a mixture of all the source signals, including the cosmological signal, a small amount of cosmological signal can leak into the reconstructed ICs, strongly biasing and reducing the amplitude of the derived cosmological signal; (2) the angular power spectrum of the dominant foreground component, Galactic synchrotron emission, has the largest amplitude for small $\ell$ (see Fig.~\ref{fig:cl}), resulting in a strong contamination in the cosmological signal at large scales in the transverse direction while increasing the risk of leakage; (3) the performance of the ICs-based reconstruction also depends on the number of measured Fourier modes in each $k$-bin, and, in the simulated data the lowest and the second lowest $k$-bins contain about 500 and 50 times fewer modes than those in the high $k$-bins, respectively.

As discussed before,  using the LRT, the three-component foreground model used in HIEMICA gives rise to the successful recovery of the HI power spectrum. In order to evaluate the recovery performance, we present the derived power spectrum in Fig.~\ref{fig:pk} in comparison with that from the existing commonly-used PCA technique. We can see that, after applying PCA to subtract the three dominant foreground components from the data cube, the residual apparently underestimates the HI power spectrum at large scales, which seems to be an over subtraction of the foregrounds. Compared with the input ones, the amplitudes in the first two $k$-bins are smaller by factors of 14.1 and 3.7 if S/N=1, and by factors of 16.1 and 4.1 if S/N=5, respectively, whereas the ratios of HIEMICA-derived power spectra and the input ones in those two bins are only about $91-92\%$ if S/N=1 and $91-94\%$ if S/N=5, respectively. In addition, the recovered amplitudes in those two $k$-bins are consistent with the input values within 2-$\sigma$.   The HIEMICA method thus does a much better job at foreground removal and cosmological signal recovery at large scales. 

More promisingly, we find that the HIEMICA approach provides an unbiased estimate of HI power spectrum for the high-$k$ modes with $k\gtrsim0.1$ Mpc$^{-1}$ where S/N tends to be low, whereas the power spectrum of the PCA-derived residual in the noise-dominant regime seems to deviate strongly from the true signal, resulting in an overestimate of HI measurements.  

Since the isotropic HI signal has the same statistical property as that of the assumed instrumental noise, commonly-used ICA-based approaches have trouble breaking the degeneracy, leaving residuals consisting of the reconstructed 21-cm signal, noise and fitting errors. As proposed in~\cite{2012MNRAS.423.2518C,2014arXiv1409.5300B}, after foreground cleaning by CCA or FASTICA, one has to  manually subtract the noise power spectrum from that of the residual data cube and obtain a comparable amplitude with the true signal. However, those previous studies show that such a scheme can cause an overfitting problem and lead to some negative or zero-closed amplitudes of the residual power spectrum in noise-dominated $k$-modes, implying that a leakage of HI signal into the subtracted foreground components is likely to be serious in regions with low S/N. Some other schemes such as weighting maps with inverse noise variance to build the frequency covariance used in PCA~\citep{2014arXiv1409.8667A} could mitigate such noise contamination.

In contrast to the above-mentioned standard PCA- and ICA-based approaches, the proposed HIEMICA method in fact provides an unbiased and efficient estimate of Gaussian components, derived from the maximum likelihood principle. If the noise covariance matrix ${\bf N}$ is exactly known, this method can optimally estimate the underlying 21-cm signal and foreground covariance matrices.  It  avoids any artificial effects that can occur when manually subtracting the noise power spectrum from the residuals after the foreground cleaning processes. Our simulations do confirm this, showing that the ensemble averaged recovered HI power spectrum converges to the true value even in low S/N regions where the noise is about an order of magnitude higher.

We note that our semi-blind approach requires exact knowledge of the noise in order to infer the 21-cm power spectrum accurately;  an incorrect noise covariance matrix can bias the signal estimate. However, a relatively lower accuracy in modelling the noise covariance matrix may be sufficient for the power spectrum estimation if the uncertainty in the noise level is not greater than the signal. In general, the corresponding impacts on the estimated parameters can be evaluated by a Monte-Carlo approach. But qualitatively, the level of the uncertainty in the estimated 21-cm power spectrum is about of the same order of magnitude as that of the noise. A simple way to understand this is to consider that in the absence of instrumental effects, if the noise is assumed to be uncorrelated, the value of $P_{\rm HI}$ which maximizes the likelihood function can be approximately estimated by $P_{\rm HI} \approx P_{\rm D} -P_{\rm F}-P_{\rm N}$ where $P_{\rm D}$, $P_{\rm F}$ and $P_{\rm N}$ are the power spectra of the data points, the foreground and noise, respectively.   As the total amplitude of the foreground fluctuations is significantly larger than that of the noise fluctuations, one can expect that the estimate of the foreground power spectrum roughly remains unchanged when slightly varying the input noise amplitude. Thus assuming an input noise with an amplitude either greater or smaller than the true one would lead to either an underestimate or an overestimate of the 21-cm power spectrum, i.e., $\Delta P_{\rm HI} \approx -\Delta P_{\rm N}$. In order to check the above estimate, we test the effect of using an input noise power spectrum with $10-50\%$ larger or smaller amplitudes than the true one and find that the resulting changes $\Delta P_{\rm HI}$ are indeed consistent with our expected variations ($-\Delta P_{\rm N}$) within a factor of 2 over all scales. In addition, incorrect noise estimates would also bias the estimate for the PCA method, since the debiasing Monte Carlo simulations would give the wrong results if the noise level were misspecified. 

In more realistic cases, an experiment may exhibit a low-level of cross-correlated Gaussian or non-Gaussian noise, both of which are unknown. In the former case, similar to the above qualitative estimation, we expect that the small level of correlation will slightly bias the signal estimate but not significantly affect it. This is because, as mentioned above, (1) the HI power spectrum is approximately determined by the difference among the diagonal components of the data point, the foreground, and noise model, not by their off-diagonal ones, and (2) the estimate of the foreground is unchanged or insensitive to the noise model. In the latter case, as the variance of the non-Gaussian noise is non-negative, the HI power spectrum is thus likely to be slightly overestimated. We will investigate the impact of noise misspecification in a future work.
 

Also, this blind component separation algorithm makes no assumptions about the spectral information of the foreground and can naturally include all instrumental effects so as to be able to coherently estimate all source components and the associated errors. More detailed comparisons between HIEMICA and other approaches for data in the presence of realistic instrumental effects will be made in future work.

\section{Conclusions}\label{sect:con}
 In this paper we present a non-parametric source separation algorithm, HIEMICA,
 which is an extension to the 3D version of the SMICA method proposed by~\cite{Snoussi:2001bw,2003MNRAS.346.1089D} for source identification in noisy mixtures of CMB maps. The HIEMICA algorithm is a fully Bayesian framework to infer the 3D power spectrum and maps of the underlying HI signal and the spatial power spectra and the frequency dependence (i.e. mixing matrix) of uncorrelated foreground components. We adapt the EM algorithm to efficiently maximize the likelihood function for unknown parameter estimates. As the statistical properties of the cosmological signal are significantly different from those of each astrophysical source, the spectral matching method in principle can blindly separate the HI signal from highly foreground-contaminated maps.

 The simulations show that HIEMICA is able to successfully reconstruct the HI 3D power spectrum across all scales and is much more robust than the PCA approach, which removes some of the 21-cm signal at large scales ($k\lesssim0.1$ Mpc$^{-1}$) and overestimates the HI power spectrum in the noise-dominated regions where noise leaks into the estimated signal. To evaluate the impact of the number of ICs in signal recovery we adapt the LRT method to rigorously assess the likelihood fit so as to determine the optimal number of foreground components used in the HIEMICA analysis. We find for the number of ICs higher than 3, a number comparable to that of the simulated astrophysical components, the reconstruction does not gain significant improvements and converges towards a stable result consistent with the input HI power spectrum.     

Although our results are quite promising, the simulated observations in this study are idealized, with no instrumental effects.  As instrument effects such as the frequency-dependent $uv$ sampling and primary beam can significantly complicate the foreground removal and bias the signal estimate, it is important to test HIEMICA in more detail in the context of specific experiment set-ups in future work.

\section*{Acknowledgments}
We acknowledge use of the open-source PETSc library~\citep{petsc-user-ref,petsc-web-page,petsc-efficient} and FFTW~\citep{FFTW05}. LZ would like to thank Flavien Vansyngel for useful comments on early drafts. LZ and PT acknowledge support from NSF awards IIS 1250720 (`Big Data' program) and AST 1216525. PMS is supported by the INFN IS PD51 ``Indark''. We would like to thank an anonymous referee for valuable suggestions which helped us to significantly improve this paper.

\bibliography{le}

\begin{thebibliography}{91}
\expandafter\ifx\csname natexlab\endcsname\relax\def\natexlab#1{#1}\fi

\bibitem[{{Alonso} {et~al.}(2015){Alonso}, {Bull}, {Ferreira}, \&
  {Santos}}]{2014arXiv1409.8667A}
{Alonso}, D., {Bull}, P., {Ferreira}, P.~G., \& {Santos}, M.~G. 2015, \mnras,
  447, 400

\bibitem[{Anderson {et~al.}(2013)Anderson, Aubourg, Bailey, Bizyaev, Blanton,
  {et~al.}}]{Anderson:2012sa}
Anderson, L., Aubourg, E., Bailey, S., Bizyaev, D., Blanton, M., {et~al.} 2013,
  Mon.Not.Roy.Astron.Soc., 427, 3435

\bibitem[{{Ansari} {et~al.}(2012{\natexlab{a}}){Ansari}, {Campagne}, {Colom},
  {Le Goff}, {Magneville}, {Martin}, {Moniez}, {Rich}, \&
  {Y{\`e}che}}]{2012A&A...540A.129A}
{Ansari}, R., {Campagne}, J.~E., {Colom}, P., {Le Goff}, J.~M., {Magneville},
  C., {Martin}, J.~M., {Moniez}, M., {Rich}, J., \& {Y{\`e}che}, C.
  2012{\natexlab{a}}, \aap, 540, A129

\bibitem[{{Ansari} {et~al.}(2012{\natexlab{b}}){Ansari}, {Campagne}, {Colom},
  {Magneville}, {Martin}, {Moniez}, {Rich}, \&
  {Y{\`e}che}}]{2012CRPhy..13...46A}
{Ansari}, R., {Campagne}, J.-E., {Colom}, P., {Magneville}, C., {Martin},
  J.-M., {Moniez}, M., {Rich}, J., \& {Y{\`e}che}, C. 2012{\natexlab{b}},
  Comptes Rendus Physique, 13, 46

\bibitem[{Balay {et~al.}(2014{\natexlab{a}})Balay, Abhyankar, Adams, Brown,
  Brune, Buschelman, Eijkhout, Gropp, Kaushik, Knepley, McInnes, Rupp, Smith,
  \& Zhang}]{petsc-user-ref}
Balay, S., Abhyankar, S., Adams, M.~F., Brown, J., Brune, P., Buschelman, K.,
  Eijkhout, V., Gropp, W.~D., Kaushik, D., Knepley, M.~G., McInnes, L.~C.,
  Rupp, K., Smith, B.~F., \& Zhang, H. 2014{\natexlab{a}}, {PETS}c Users
  Manual, Tech. Rep. ANL-95/11 - Revision 3.5, Argonne National Laboratory

\bibitem[{Balay {et~al.}(2014{\natexlab{b}})Balay, Abhyankar, Adams, Brown,
  Brune, Buschelman, Eijkhout, Gropp, Kaushik, Knepley, McInnes, Rupp, Smith,
  \& Zhang}]{petsc-web-page}
---. 2014{\natexlab{b}}, {PETS}c {W}eb page, \url{http://www.mcs.anl.gov/petsc}

\bibitem[{Balay {et~al.}(1997)Balay, Gropp, McInnes, \&
  Smith}]{petsc-efficient}
Balay, S., Gropp, W.~D., McInnes, L.~C., \& Smith, B.~F. 1997, in Modern
  Software Tools in Scientific Computing, ed. E.~Arge, A.~M. Bruaset, \& H.~P.
  Langtangen (Birkh{\"{a}}user Press), 163--202

\bibitem[{{Bonaldi} \& {Brown}(2015)}]{2014arXiv1409.5300B}
{Bonaldi}, A. \& {Brown}, M.~L. 2015, \mnras, 447, 1973

\bibitem[{{Bowman} {et~al.}(2006){Bowman}, {Morales}, \&
  {Hewitt}}]{2006ApJ...638...20B}
{Bowman}, J.~D., {Morales}, M.~F., \& {Hewitt}, J.~N. 2006, \apj, 638, 20

\bibitem[{Bowman {et~al.}(2009)Bowman, Morales, \&
  Hewitt}]{2009ApJ...695..183B}
Bowman, J.~D., Morales, M.~F., \& Hewitt, J.~N. 2009, Astrophys.J., 695, 183

\bibitem[{{Bull} {et~al.}(2015){Bull}, {Ferreira}, {Patel}, \&
  {Santos}}]{2014arXiv1405.1452B}
{Bull}, P., {Ferreira}, P.~G., {Patel}, P., \& {Santos}, M.~G. 2015, \apj, 803,
  21

\bibitem[{{Cardoso} {et~al.}(2008){Cardoso}, {Martin}, {Delabrouille},
  {Betoule}, \& {Patanchon}}]{2008arXiv0803.1814C}
{Cardoso}, J.-F., {Martin}, M., {Delabrouille}, J., {Betoule}, M., \&
  {Patanchon}, G. 2008, IEEE Journal of Selected Topics in Signal Processing,
  2, 735

\bibitem[{{Chang} {et~al.}(2010){Chang}, {Pen}, {Bandura}, \&
  {Peterson}}]{2010Natur.466..463C}
{Chang}, T., {Pen}, U., {Bandura}, K., \& {Peterson}, J.~B. 2010, \nat, 466,
  463

\bibitem[{{Chang} {et~al.}(2008){Chang}, {Pen}, {Peterson}, \&
  {McDonald}}]{Chang2008}
{Chang}, T.-C., {Pen}, U.-L., {Peterson}, J.~B., \& {McDonald}, P. 2008,
  Physical Review Letters, 100, 091303

\bibitem[{{Chapman} {et~al.}(2012){Chapman}, {Abdalla}, {Harker}, {Jeli{\'c}},
  {Labropoulos}, {Zaroubi}, {Brentjens}, {de Bruyn}, \&
  {Koopmans}}]{2012MNRAS.423.2518C}
{Chapman}, E., {Abdalla}, F.~B., {Harker}, G., {Jeli{\'c}}, V., {Labropoulos},
  P., {Zaroubi}, S., {Brentjens}, M.~A., {de Bruyn}, A.~G., \& {Koopmans},
  L.~V.~E. 2012, \mnras, 423, 2518

\bibitem[{{Chen}(2012)}]{2012IJMPS..12..256C}
{Chen}, X. 2012, International Journal of Modern Physics Conference Series, 12,
  256

\bibitem[{Chen \& Miralda-Escude(2004)}]{Chen:2003gc}
Chen, X.-L. \& Miralda-Escude, J. 2004, Astrophys.J., 602, 1

\bibitem[{{Colless} {et~al.}(2003){Colless}, {Peterson}, {Jackson}, {Peacock},
  {Cole}, {Norberg}, {Baldry}, {Baugh}, {Bland-Hawthorn}, {Bridges}, {Cannon},
  {Collins}, {Couch}, {Cross}, {Dalton}, {De Propris}, {Driver}, {Efstathiou},
  {Ellis}, {Frenk}, {Glazebrook}, {Lahav}, {Lewis}, {Lumsden}, {Maddox},
  {Madgwick}, {Sutherland}, \& {Taylor}}]{2003astro.ph..6581C}
{Colless}, M., {Peterson}, B.~A., {Jackson}, C., {Peacock}, J.~A., {Cole}, S.,
  {Norberg}, P., {Baldry}, I.~K., {Baugh}, C.~M., {Bland-Hawthorn}, J.,
  {Bridges}, T., {Cannon}, R., {Collins}, C., {Couch}, W., {Cross}, N.,
  {Dalton}, G., {De Propris}, R., {Driver}, S.~P., {Efstathiou}, G., {Ellis},
  R.~S., {Frenk}, C.~S., {Glazebrook}, K., {Lahav}, O., {Lewis}, I., {Lumsden},
  S., {Maddox}, S., {Madgwick}, D., {Sutherland}, W., \& {Taylor}, K. 2003,
  ArXiv Astrophysics e-prints

\bibitem[{{Cooray} \& {Furlanetto}(2004)}]{2004ApJ...606L...5C}
{Cooray}, A. \& {Furlanetto}, S.~R. 2004, \apjl, 606, L5

\bibitem[{{Datta} {et~al.}(2010){Datta}, {Bowman}, \&
  {Carilli}}]{2010ApJ...724..526D}
{Datta}, A., {Bowman}, J.~D., \& {Carilli}, C.~L. 2010, \apj, 724, 526

\bibitem[{{Davies} {et~al.}(2006){Davies}, {Dickinson}, {Banday}, {Jaffe},
  {G{\'o}rski}, \& {Davis}}]{2006MNRAS.370.1125D}
{Davies}, R.~D., {Dickinson}, C., {Banday}, A.~J., {Jaffe}, T.~R.,
  {G{\'o}rski}, K.~M., \& {Davis}, R.~J. 2006, \mnras, 370, 1125

\bibitem[{{de Oliveira-Costa} {et~al.}(2008){de Oliveira-Costa}, {Tegmark},
  {Gaensler}, {Jonas}, {Landecker}, \& {Reich}}]{2008MNRAS.388..247D}
{de Oliveira-Costa}, A., {Tegmark}, M., {Gaensler}, B.~M., {Jonas}, J.,
  {Landecker}, T.~L., \& {Reich}, P. 2008, \mnras, 388, 247

\bibitem[{{Delabrouille} {et~al.}(2003){Delabrouille}, {Cardoso}, \&
  {Patanchon}}]{2003MNRAS.346.1089D}
{Delabrouille}, J., {Cardoso}, J.-F., \& {Patanchon}, G. 2003, \mnras, 346,
  1089

\bibitem[{{Di Matteo} {et~al.}(2004){Di Matteo}, {Ciardi}, \&
  {Miniati}}]{2004MNRAS.355.1053D}
{Di Matteo}, T., {Ciardi}, B., \& {Miniati}, F. 2004, \mnras, 355, 1053

\bibitem[{{Di Matteo} {et~al.}(2002){Di Matteo}, {Perna}, {Abel}, \&
  {Rees}}]{2002ApJ...564..576D}
{Di Matteo}, T., {Perna}, R., {Abel}, T., \& {Rees}, M.~J. 2002, \apj, 564, 576

\bibitem[{{Dillon} {et~al.}(2013){Dillon}, {Liu}, \&
  {Tegmark}}]{2013PhRvD..87d3005D}
{Dillon}, J.~S., {Liu}, A., \& {Tegmark}, M. 2013, \prd, 87, 043005

\bibitem[{Drinkwater {et~al.}(2010)Drinkwater, Jurek, Blake, Woods, Pimbblet,
  {et~al.}}]{2010MNRAS.401.1429D}
Drinkwater, M.~J., Jurek, R.~J., Blake, C., Woods, D., Pimbblet, K.~A.,
  {et~al.} 2010, Mon.Not.Roy.Astron.Soc., 401, 1429

\bibitem[{{Eisenstein} \& {Hu}(1998)}]{1998ApJ...496..605E}
{Eisenstein}, D.~J. \& {Hu}, W. 1998, \apj, 496, 605

\bibitem[{Fessler \& Sutton(2003)}]{Fessler03nonuniformfast}
Fessler, J.~A. \& Sutton, B.~P. 2003, IEEE Trans. Signal Process, 51, 560

\bibitem[{Frigo \& Johnson(2005)}]{FFTW05}
Frigo, M. \& Johnson, S.~G. 2005, Proceedings of the IEEE, 93, 216, special
  issue on ``Program Generation, Optimization, and Platform Adaptation''

\bibitem[{Furlanetto {et~al.}(2006)Furlanetto, Oh, \&
  Briggs}]{Furlanetto:2006jb}
Furlanetto, S., Oh, S.~P., \& Briggs, F. 2006, Phys.Rept., 433, 181

\bibitem[{{Gleser} {et~al.}(2008){Gleser}, {Nusser}, \&
  {Benson}}]{2008MNRAS.391..383G}
{Gleser}, L., {Nusser}, A., \& {Benson}, A.~J. 2008, \mnras, 391, 383

\bibitem[{{Harker} {et~al.}(2010){Harker}, {Zaroubi}, {Bernardi}, {Brentjens},
  {de Bruyn}, {Ciardi}, {Jeli{\'c}}, {Koopmans}, {Labropoulos}, {Mellema},
  {Offringa}, {Pandey}, {Pawlik}, {Schaye}, {Thomas}, \&
  {Yatawatta}}]{2010MNRAS.405.2492H}
{Harker}, G., {Zaroubi}, S., {Bernardi}, G., {Brentjens}, M.~A., {de Bruyn},
  A.~G., {Ciardi}, B., {Jeli{\'c}}, V., {Koopmans}, L.~V.~E., {Labropoulos},
  P., {Mellema}, G., {Offringa}, A., {Pandey}, V.~N., {Pawlik}, A.~H.,
  {Schaye}, J., {Thomas}, R.~M., \& {Yatawatta}, S. 2010, \mnras, 405, 2492

\bibitem[{{Harker} {et~al.}(2009{\natexlab{a}}){Harker}, {Zaroubi}, {Bernardi},
  {Brentjens}, {de Bruyn}, {Ciardi}, {Jeli{\'c}}, {Koopmans}, {Labropoulos},
  {Mellema}, {Offringa}, {Pandey}, {Schaye}, {Thomas}, \&
  {Yatawatta}}]{2009MNRAS.397.1138H}
{Harker}, G., {Zaroubi}, S., {Bernardi}, G., {Brentjens}, M.~A., {de Bruyn},
  A.~G., {Ciardi}, B., {Jeli{\'c}}, V., {Koopmans}, L.~V.~E., {Labropoulos},
  P., {Mellema}, G., {Offringa}, A., {Pandey}, V.~N., {Schaye}, J., {Thomas},
  R.~M., \& {Yatawatta}, S. 2009{\natexlab{a}}, \mnras, 397, 1138

\bibitem[{{Harker} {et~al.}(2009{\natexlab{b}}){Harker}, {Zaroubi}, {Thomas},
  {Jeli{\'c}}, {Labropoulos}, {Mellema}, {Iliev}, {Bernardi}, {Brentjens}, {de
  Bruyn}, {Ciardi}, {Koopmans}, {Pandey}, {Pawlik}, {Schaye}, \&
  {Yatawatta}}]{2009MNRAS.393.1449H}
{Harker}, G.~J.~A., {Zaroubi}, S., {Thomas}, R.~M., {Jeli{\'c}}, V.,
  {Labropoulos}, P., {Mellema}, G., {Iliev}, I.~T., {Bernardi}, G.,
  {Brentjens}, M.~A., {de Bruyn}, A.~G., {Ciardi}, B., {Koopmans}, L.~V.~E.,
  {Pandey}, V.~N., {Pawlik}, A.~H., {Schaye}, J., \& {Yatawatta}, S.
  2009{\natexlab{b}}, \mnras, 393, 1449

\bibitem[{{Hinshaw} {et~al.}(2013){Hinshaw}, {Larson}, {Komatsu},
  {et~al.}}]{2013ApJS..208...19H}
{Hinshaw}, G., {Larson}, D., {Komatsu}, E., {et~al.} 2013, \apjs, 208, 19

\bibitem[{Hyv\"{a}rinen(1999)}]{fastica99}
Hyv\"{a}rinen, A. 1999, IEEE Trans. On Neural Networks, 10, 626

\bibitem[{Hyv\"{a}rinen \& Oja(2000)}]{fastica10}
Hyv\"{a}rinen, A. \& Oja, E. 2000, Neural Netw., 13, 411

\bibitem[{Jeffreys(1961)}]{Jeffreys:1961}
Jeffreys, H. 1961, Theory of Probability (3rd Edition) (New York: Oxford
  University Press)

\bibitem[{{Jeli{\'c}} {et~al.}(2008){Jeli{\'c}}, {Zaroubi}, {Labropoulos},
  {Thomas}, {Bernardi}, {Brentjens}, {de Bruyn}, {Ciardi}, {Harker},
  {Koopmans}, {Pandey}, {Schaye}, \& {Yatawatta}}]{2008MNRAS.389.1319J}
{Jeli{\'c}}, V., {Zaroubi}, S., {Labropoulos}, P., {Thomas}, R.~M., {Bernardi},
  G., {Brentjens}, M.~A., {de Bruyn}, A.~G., {Ciardi}, B., {Harker}, G.,
  {Koopmans}, L.~V.~E., {Pandey}, V.~N., {Schaye}, J., \& {Yatawatta}, S. 2008,
  \mnras, 389, 1319

\bibitem[{{Jewell} {et~al.}(2004){Jewell}, {Levin}, \&
  {Anderson}}]{2004ApJ...609....1J}
{Jewell}, J., {Levin}, S., \& {Anderson}, C.~H. 2004, \apj, 609, 1

\bibitem[{{Kamionkowski} {et~al.}(1997){Kamionkowski}, {Kosowsky}, \&
  {Stebbins}}]{kkslett}
{Kamionkowski}, M., {Kosowsky}, A., \& {Stebbins}, A. 1997, Physical Review
  Letters, 78, 2058

\bibitem[{{Kinney} {et~al.}(2006){Kinney}, {Kolb}, {Melchiorri}, \&
  {Riotto}}]{2006PhRvD..74b3502K}
{Kinney}, W.~H., {Kolb}, E.~W., {Melchiorri}, A., \& {Riotto}, A. 2006, \prd,
  74, 023502

\bibitem[{{Liu} {et~al.}(2014{\natexlab{a}}){Liu}, {Parsons}, \&
  {Trott}}]{2014PhRvD..90b3018L}
{Liu}, A., {Parsons}, A.~R., \& {Trott}, C.~M. 2014{\natexlab{a}}, \prd, 90,
  023018

\bibitem[{{Liu} {et~al.}(2014{\natexlab{b}}){Liu}, {Parsons}, \&
  {Trott}}]{2014PhRvD..90b3019L}
---. 2014{\natexlab{b}}, \prd, 90, 023019

\bibitem[{{Liu} \& {Tegmark}(2012)}]{2012MNRAS.419.3491L}
{Liu}, A. \& {Tegmark}, M. 2012, \mnras, 419, 3491

\bibitem[{{Liu} {et~al.}(2009{\natexlab{a}}){Liu}, {Tegmark}, {Bowman},
  {Hewitt}, \& {Zaldarriaga}}]{2009MNRAS.398..401L}
{Liu}, A., {Tegmark}, M., {Bowman}, J., {Hewitt}, J., \& {Zaldarriaga}, M.
  2009{\natexlab{a}}, \mnras, 398, 401

\bibitem[{{Liu} {et~al.}(2009{\natexlab{b}}){Liu}, {Tegmark}, \&
  {Zaldarriaga}}]{2009MNRAS.394.1575L}
{Liu}, A., {Tegmark}, M., \& {Zaldarriaga}, M. 2009{\natexlab{b}}, \mnras, 394,
  1575

\bibitem[{{Loeb} \& {Wyithe}(2008)}]{2008PhRvL.100p1301L}
{Loeb}, A. \& {Wyithe}, J.~S.~B. 2008, Physical Review Letters, 100, 161301

\bibitem[{{Loeb} \& {Zaldarriaga}(2004)}]{2004PhRvL..92u1301L}
{Loeb}, A. \& {Zaldarriaga}, M. 2004, Physical Review Letters, 92, 211301

\bibitem[{Madau {et~al.}(1997)Madau, Meiksin, \& Rees}]{Madau:1996cs}
Madau, P., Meiksin, A., \& Rees, M.~J. 1997, Astrophys.J., 475, 429

\bibitem[{Maldacena(2003)}]{Maldacena:2002vr}
Maldacena, J.~M. 2003, JHEP, 0305, 013

\bibitem[{Mao {et~al.}(2008)Mao, Tegmark, McQuinn, Zaldarriaga, \&
  Zahn}]{Mao:2008ug}
Mao, Y., Tegmark, M., McQuinn, M., Zaldarriaga, M., \& Zahn, O. 2008,
  Phys.Rev., D78, 023529

\bibitem[{{Masui} {et~al.}(2013){Masui}, {Switzer}, {Banavar}, {Bandura},
  {Blake}, {Calin}, {Chang}, {Chen}, {Li}, {Liao}, {Natarajan}, {Pen},
  {Peterson}, {Shaw}, \& {Voytek}}]{Masui2012}
{Masui}, K.~W., {Switzer}, E.~R., {Banavar}, N., {Bandura}, K., {Blake}, C.,
  {Calin}, L.-M., {Chang}, T.-C., {Chen}, X., {Li}, Y.-C., {Liao}, Y.-W.,
  {Natarajan}, A., {Pen}, U.-L., {Peterson}, J.~B., {Shaw}, J.~R., \& {Voytek},
  T.~C. 2013, \apjl, 763, L20

\bibitem[{{McQuinn} {et~al.}(2006){McQuinn}, {Zahn}, {Zaldarriaga},
  {Hernquist}, \& {Furlanetto}}]{2006ApJ...653..815M}
{McQuinn}, M., {Zahn}, O., {Zaldarriaga}, M., {Hernquist}, L., \& {Furlanetto},
  S.~R. 2006, \apj, 653, 815

\bibitem[{{Morales} {et~al.}(2006){Morales}, {Bowman}, \&
  {Hewitt}}]{2006ApJ...648..767M}
{Morales}, M.~F., {Bowman}, J.~D., \& {Hewitt}, J.~N. 2006, \apj, 648, 767

\bibitem[{{Morales} {et~al.}(2012){Morales}, {Hazelton}, {Sullivan}, \&
  {Beardsley}}]{2012ApJ...752..137M}
{Morales}, M.~F., {Hazelton}, B., {Sullivan}, I., \& {Beardsley}, A. 2012,
  \apj, 752, 137

\bibitem[{{Morales} \& {Hewitt}(2004)}]{2004ApJ...615....7M}
{Morales}, M.~F. \& {Hewitt}, J. 2004, \apj, 615, 7

\bibitem[{{Morales} \& {Wyithe}(2010)}]{2010ARA&A..48..127M}
{Morales}, M.~F. \& {Wyithe}, J.~S.~B. 2010, Annual Review of Astronomy and
  Astrophysics, 48, 127

\bibitem[{{Nan} {et~al.}(2011){Nan}, {Li}, {Jin}, {Wang}, {Zhu}, {Zhu},
  {Zhang}, {Yue}, \& {Qian}}]{2011IJMPD..20..989N}
{Nan}, R., {Li}, D., {Jin}, C., {Wang}, Q., {Zhu}, L., {Zhu}, W., {Zhang}, H.,
  {Yue}, Y., \& {Qian}, L. 2011, International Journal of Modern Physics D, 20,
  989

\bibitem[{{Oh} \& {Mack}(2003)}]{2003MNRAS.346..871O}
{Oh}, S.~P. \& {Mack}, K.~J. 2003, \mnras, 346, 871

\bibitem[{{Parsons} {et~al.}(2012){Parsons}, {Pober}, {Aguirre}, {Carilli},
  {Jacobs}, \& {Moore}}]{2012ApJ...756..165P}
{Parsons}, A.~R., {Pober}, J.~C., {Aguirre}, J.~E., {Carilli}, C.~L., {Jacobs},
  D.~C., \& {Moore}, D.~F. 2012, \apj, 756, 165

\bibitem[{{Peiris} {et~al.}(2003){Peiris}, {Komatsu}, {Verde}, {Spergel},
  {Bennett}, {Halpern}, {Hinshaw}, {Jarosik}, {Kogut}, {Limon}, {Meyer},
  {Page}, {Tucker}, {Wollack}, \& {Wright}}]{2003ApJS..148..213P}
{Peiris}, H.~V., {Komatsu}, E., {Verde}, L., {Spergel}, D.~N., {Bennett},
  C.~L., {Halpern}, M., {Hinshaw}, G., {Jarosik}, N., {Kogut}, A., {Limon}, M.,
  {Meyer}, S.~S., {Page}, L., {Tucker}, G.~S., {Wollack}, E., \& {Wright},
  E.~L. 2003, \apjs, 148, 213

\bibitem[{{Pen} {et~al.}(2009){Pen}, {Staveley-Smith}, {Peterson}, \&
  {Chang}}]{2009MNRAS.394L...6P}
{Pen}, U.-L., {Staveley-Smith}, L., {Peterson}, J.~B., \& {Chang}, T.-C. 2009,
  \mnras, 394, L6

\bibitem[{Peterson {et~al.}(2006)Peterson, Bandura, \&
  Pen}]{2006astro.ph..6104P}
Peterson, J.~B., Bandura, K., \& Pen, U.~L. 2006, in {Proceedings, 41st
  Recontres de Moriond, 2006 Contents and structure of the universe}, 283--289

\bibitem[{{Petrovic} \& {Oh}(2011)}]{2011MNRAS.413.2103P}
{Petrovic}, N. \& {Oh}, S.~P. 2011, \mnras, 413, 2103

\bibitem[{{Planck Collaboration} {et~al.}(2014{\natexlab{a}}){Planck
  Collaboration}, {Ade}, {Aghanim}, {Armitage-Caplan}, {Arnaud}, {Ashdown},
  {Atrio-Barandela}, {Aumont}, {Baccigalupi}, {Banday}, \&
  et~al.}]{2014A&A...571A..12P}
{Planck Collaboration}, {Ade}, P.~A.~R., {Aghanim}, N., {Armitage-Caplan}, C.,
  {Arnaud}, M., {Ashdown}, M., {Atrio-Barandela}, F., {Aumont}, J.,
  {Baccigalupi}, C., {Banday}, A.~J., \& et~al. 2014{\natexlab{a}}, \aap, 571,
  A12

\bibitem[{{Planck Collaboration} {et~al.}(2014{\natexlab{b}}){Planck
  Collaboration}, {Ade}, {Aghanim}, {et~al.}}]{2014A&A...571A..16P}
{Planck Collaboration}, {Ade}, P.~A.~R., {Aghanim}, N., {et~al.}
  2014{\natexlab{b}}, \aap, 571, A16

\bibitem[{{Pober} {et~al.}(2014){Pober}, {Liu}, {Dillon}, {Aguirre}, {Bowman},
  {Bradley}, {Carilli}, {DeBoer}, {Hewitt}, {Jacobs}, {McQuinn}, {Morales},
  {Parsons}, {Tegmark}, \& {Werthimer}}]{2014ApJ...782...66P}
{Pober}, J.~C., {Liu}, A., {Dillon}, J.~S., {Aguirre}, J.~E., {Bowman}, J.~D.,
  {Bradley}, R.~F., {Carilli}, C.~L., {DeBoer}, D.~R., {Hewitt}, J.~N.,
  {Jacobs}, D.~C., {McQuinn}, M., {Morales}, M.~F., {Parsons}, A.~R.,
  {Tegmark}, M., \& {Werthimer}, D.~J. 2014, \apj, 782, 66

\bibitem[{{Pober} {et~al.}(2013{\natexlab{a}}){Pober}, {Parsons}, {Aguirre},
  {Ali}, {Bradley}, {Carilli}, {DeBoer}, {Dexter}, {Gugliucci}, {Jacobs},
  {Klima}, {MacMahon}, {Manley}, {Moore}, {Stefan}, \&
  {Walbrugh}}]{2013ApJ...768L..36P}
{Pober}, J.~C., {Parsons}, A.~R., {Aguirre}, J.~E., {Ali}, Z., {Bradley},
  R.~F., {Carilli}, C.~L., {DeBoer}, D., {Dexter}, M., {Gugliucci}, N.~E.,
  {Jacobs}, D.~C., {Klima}, P.~J., {MacMahon}, D., {Manley}, J., {Moore},
  D.~F., {Stefan}, I.~I., \& {Walbrugh}, W.~P. 2013{\natexlab{a}}, \apjl, 768,
  L36

\bibitem[{{Pober} {et~al.}(2013{\natexlab{b}}){Pober}, {Parsons}, {DeBoer},
  {McDonald}, {McQuinn}, {Aguirre}, {Ali}, {Bradley}, {Chang}, \&
  {Morales}}]{2013AJ....145...65P}
{Pober}, J.~C., {Parsons}, A.~R., {DeBoer}, D.~R., {McDonald}, P., {McQuinn},
  M., {Aguirre}, J.~E., {Ali}, Z., {Bradley}, R.~F., {Chang}, T.-C., \&
  {Morales}, M.~F. 2013{\natexlab{b}}, \aj, 145, 65

\bibitem[{{Pritchard} \& {Loeb}(2012)}]{2012RPPh...75h6901P}
{Pritchard}, J.~R. \& {Loeb}, A. 2012, Reports on Progress in Physics, 75,
  086901

\bibitem[{{Santos} {et~al.}(2005){Santos}, {Cooray}, \&
  {Knox}}]{2005ApJ...625..575S}
{Santos}, M.~G., {Cooray}, A., \& {Knox}, L. 2005, \apj, 625, 575

\bibitem[{{Seo} \& {Eisenstein}(2003)}]{2003ApJ...598..720S}
{Seo}, H.-J. \& {Eisenstein}, D.~J. 2003, \apj, 598, 720

\bibitem[{{Shaw} {et~al.}(2014){Shaw}, {Sigurdson}, {Pen}, {Stebbins}, \&
  {Sitwell}}]{2014ApJ...781...57S}
{Shaw}, J.~R., {Sigurdson}, K., {Pen}, U.-L., {Stebbins}, A., \& {Sitwell}, M.
  2014, \apj, 781, 57

\bibitem[{{Smoot} {et~al.}(1992){Smoot}, {Bennett}, {Kogut},
  {et~al.}}]{smootetal1992}
{Smoot}, G.~F., {Bennett}, C.~L., {Kogut}, A., {et~al.} 1992, \apjl, 396, L1

\bibitem[{{Snoussi} {et~al.}(2002){Snoussi}, {Patanchon},
  {Mac{\'{\i}}as-P{\'e}rez}, {Mohammad-Djafari}, \&
  {Delabrouille}}]{Snoussi:2001bw}
{Snoussi}, H., {Patanchon}, G., {Mac{\'{\i}}as-P{\'e}rez}, J.~F.,
  {Mohammad-Djafari}, A., \& {Delabrouille}, J. 2002, in American Institute of
  Physics Conference Series, Vol. 617, Bayesian Inference and Maximum Entropy
  Methods in Science and Engineering, ed. R.~L. {Fry}, 125--140

\bibitem[{{Sutter} {et~al.}(2012){Sutter}, {Wandelt}, \&
  {Malu}}]{2012ApJS..202....9S}
{Sutter}, P.~M., {Wandelt}, B.~D., \& {Malu}, S.~S. 2012, \apjs, 202, 9

\bibitem[{{Switzer} {et~al.}(2013){Switzer}, {Masui}, {Bandura}, {Calin},
  {Chang}, {Chen}, {Li}, {Liao}, {Natarajan}, {Pen}, {Peterson}, {Shaw}, \&
  {Voytek}}]{Switzer:2013ewa}
{Switzer}, E.~R., {Masui}, K.~W., {Bandura}, K., {Calin}, L.-M., {Chang},
  T.-C., {Chen}, X.-L., {Li}, Y.-C., {Liao}, Y.-W., {Natarajan}, A., {Pen},
  U.-L., {Peterson}, J.~B., {Shaw}, J.~R., \& {Voytek}, T.~C. 2013, \mnras,
  434, L46

\bibitem[{{Trott} {et~al.}(2012){Trott}, {Wayth}, \&
  {Tingay}}]{2012ApJ...757..101T}
{Trott}, C.~M., {Wayth}, R.~B., \& {Tingay}, S.~J. 2012, \apj, 757, 101

\bibitem[{{Vansyngel} {et~al.}(2014){Vansyngel}, {Wandelt}, {Cardoso}, \&
  {Benabed}}]{2014arXiv1409.0858V}
{Vansyngel}, F., {Wandelt}, B.~D., {Cardoso}, J.-F., \& {Benabed}, K. 2014,
  ArXiv e-prints

\bibitem[{{Vedantham} {et~al.}(2012){Vedantham}, {Udaya Shankar}, \&
  {Subrahmanyan}}]{2012ApJ...745..176V}
{Vedantham}, H., {Udaya Shankar}, N., \& {Subrahmanyan}, R. 2012, \apj, 745,
  176

\bibitem[{{Wandelt} {et~al.}(2004){Wandelt}, {Larson}, \&
  {Lakshminarayanan}}]{2004PhRvD..70h3511W}
{Wandelt}, B.~D., {Larson}, D.~L., \& {Lakshminarayanan}, A. 2004, \prd, 70,
  083511

\bibitem[{{Wang} {et~al.}(2009){Wang}, {Chen}, {Zheng}, {Wu}, {Zhang}, \&
  {Zhao}}]{2009MNRAS.394.1775W}
{Wang}, X., {Chen}, X., {Zheng}, Z., {Wu}, F., {Zhang}, P., \& {Zhao}, Y. 2009,
  \mnras, 394, 1775

\bibitem[{{Wang} {et~al.}(2006){Wang}, {Tegmark}, {Santos}, \&
  {Knox}}]{2006ApJ...650..529W}
{Wang}, X., {Tegmark}, M., {Santos}, M.~G., \& {Knox}, L. 2006, \apj, 650, 529

\bibitem[{{Wolz} {et~al.}(2014){Wolz}, {Abdalla}, {Blake}, {Shaw}, {Chapman},
  \& {Rawlings}}]{2013arXiv1310.8144W}
{Wolz}, L., {Abdalla}, F.~B., {Blake}, C., {Shaw}, J.~R., {Chapman}, E., \&
  {Rawlings}, S. 2014, \mnras, 441, 3271

\bibitem[{Wyithe {et~al.}(2008)Wyithe, Loeb, \& Geil}]{wyithe08}
Wyithe, S., Loeb, A., \& Geil, P. 2008, \mnras, 383, 1195

\bibitem[{{Xu} {et~al.}(2015){Xu}, {Wang}, \& {Chen}}]{2014arXiv1410.7794X}
{Xu}, Y., {Wang}, X., \& {Chen}, X. 2015, \apj, 798, 40

\bibitem[{{York} {et~al.}(2000){York}, {Adelman}, {Anderson},
  {et~al.}}]{York:2000gk}
{York}, D.~G., {Adelman}, J., {Anderson}, Jr., J.~E., {et~al.} 2000, \aj, 120,
  1579

\bibitem[{{Zaldarriaga} {et~al.}(2004){Zaldarriaga}, {Furlanetto}, \&
  {Hernquist}}]{2004ApJ...608..622Z}
{Zaldarriaga}, M., {Furlanetto}, S.~R., \& {Hernquist}, L. 2004, \apj, 608, 622

\bibitem[{{Zaldarriaga} \& {Seljak}(1998)}]{zalsel1998}
{Zaldarriaga}, M. \& {Seljak}, U. 1998, \prd, 58, 023003

\end{thebibliography}
\bibliographystyle{apj}
\nocite{*}

\appendix

We here present a derivation of an approximate expression for the posterior distribution of the signal power spectrum in terms of the sampling distribution of the estimator so as to calculate Bayesian credible intervals of the inferred parameter. The sampling distribution of the estimator in our case can be derived from a Monte Carlo (MC) distribution of the Maximum Likelihood Estimator (MLE) of the signal power spectrum $P$.

The power spectrum is simply the variance of the Fourier mode coefficients. For simplicity, if we assume that the all the Fourier modes are independent and the variance of all the Fourier mode coefficients within the bin is the same, the posterior distribution of $P$ given data $d_i$ for $i=1,\cdots,n$ can be expressed in an ideal noise-free case as 
\begin{equation}\label{eq:lp}
p(P|d) \propto {\cal L}(d|P)p(P)\propto P^{-\frac{n}{2}}e^{-\frac{1}{2}\sum^n_{i=1}d^2_i/P} P^{-1}\,, 
\end{equation}
where the non-informative Jeffreys prior~\citep{Jeffreys:1961}, $p(P)=P^{-1}$, has been applied on the power spectrum. As we know, the MLE $\hat{P} =\frac{1}{n}\sum^n_{i=1}d^2_i$ is unbiased over data sets drawn from 
\begin{equation}
\left<\hat{P}\right> =P\,
\end{equation}
and has variance 
\begin{equation}
\left<(\Delta\hat{P})^2\right> =\frac{2}{n}P^2\,.
\end{equation}

If we want to construct an approximate Bayesian posterior from MC samples from the sampling density of the MLE $\hat{P}$ and the effective number of degrees of freedom is unknown, then we can determine it from the mean and variance of $\hat{P}$ using
\begin{equation}\label{eq:en}
\tilde{n} = 2\frac{\left<\tilde{P}\right>}{\left<(\Delta\hat{P})^2\right>}\,.
\end{equation}

Assuming that noise is subdominant, inserting Eq.~\ref{eq:en} into Eq.~\ref{eq:lp}, we can use $\hat{P}$ and the effective number of degrees $\tilde{n}$ of freedom to approximate the posterior for $P$ as
\begin{equation}
p(P|d)\appropto e^{-\frac{1}{2}\tilde{n}\hat{P}/P}/P^{\frac{\tilde{n}+2}{2}}\,,
\end{equation}
from which the asymmetric credible interval can be derived.

\end{document}